\documentclass[sn-mathphys,Numbered]{sn-jnl}%

\usepackage{graphicx}%
\usepackage{multirow}%
\usepackage{amsmath,amssymb,amsfonts}%
\usepackage{amsthm}%
\usepackage{mathrsfs}%
\usepackage[title]{appendix}%
\usepackage{xcolor}
\usepackage{textcomp}%
\usepackage{manyfoot}%
\usepackage{booktabs}%
\usepackage{algorithm}%
\usepackage{algorithmicx}%
\usepackage{algpseudocode}%
\usepackage{listings}%

\usepackage{caption}
\usepackage{subcaption}
\usepackage{amsmath}
\usepackage{amsfonts}
\usepackage{siunitx}
\usepackage{url}
\usepackage{placeins}
\DeclareSIUnit\angstrom{\text {Å}}
\DeclareSIUnit\isovalue{\ensuremath{electrons/bohr^3}}
\usepackage{gensymb}

\newcommand{\ket}[1]{ |#1 \rangle}

\definecolor{blue_A}{HTML}{3f4fa8}
\definecolor{green_B}{HTML}{218d39}
\definecolor{orange_C}{HTML}{c05627}

\newcommand{\SM}{\boldsymbol{S}}
\newcommand{\DM}{\boldsymbol{\rho}}
\newcommand{\DDM}{\boldsymbol{\Delta \rho}}
\newcommand{\CM}{\boldsymbol{C}}
\DeclareMathOperator*{\tr}{trace}

\newcommand{\colorA}[1]{\textcolor{blue_A}{#1}}
\newcommand{\colorB}[1]{\textcolor{green_B}{#1}}
\newcommand{\colorC}[1]{\textcolor{orange_C}{#1}}

\begin{document}

\title[]{Understanding X-ray absorption in liquid water:
triple excitations in multilevel coupled cluster theory}

\author[1]{Sarai Dery Folkestad}
\equalcont{These authors contributed equally to this work.}
\email{sarai.d.folkestad@ntnu.no}
\author[1]{Alexander C. Paul}
\email{alexander.c.paul@ntnu.no}
\equalcont{These authors contributed equally to this work.}
\author[1]{Regina Paul (née Matveeva)}
\email{regina.paul@ntnu.no}
\equalcont{These authors contributed equally to this work.}
\author[2]{Sonia Coriani}
\email{soco@kemi.dtu.dk}
\author[3]{Michael Odelius}
\email{odelius@fysik.su.se}
\author[4]{Marcella Iannuzzi}
\email{marcella.iannuzzi@chem.uzh.ch}
\author[1,5]{Henrik Koch}
\email{henrik.koch@ntnu.no}

\affil[1]{Department of Chemistry, Norwegian University of Science and Technology, NTNU, 7491 Trondheim, Norway}
\affil[2]{Department of Chemistry, Technical University of Denmark, DTU, 2800 Kongens Lyngby, Denmark}
\affil[3]{Department of Physics, Stockholm University, 10691 Stockholm, Sweden}
\affil[4]{Department of Chemistry, University of Zurich, 8057 Zürich, Switzerland}
\affil[5]{Scuola Normale Superiore, Piazza dei Cavaleri 7, 56126 Pisa, Italy}

\abstract{We present the first successful application of the coupled cluster
approach to simulate the X-ray absorption (XA) spectrum of liquid water.
The system size limitations of standard coupled cluster theory
are overcome by employing a newly developed coupled cluster method for large molecular systems.
This method combines coupled cluster singles, doubles, and perturbative triples
in a multilevel framework (MLCC3-in-HF) and
is able to describe the delicate nature of intermolecular interactions in liquid water.
Using molecular geometries from state-of-the-art path-integral molecular dynamics,
we obtain excellent agreement with experimental spectra.
Additionally,
we show that an accurate description of the electronic structure
within the first solvation shell is sufficient to model the XA spectrum of liquid water.
Furthermore, we present a rigorous charge transfer analysis
with unprecedented reliability, achieved through MLCC3-in-HF.
This analysis aligns with previous studies regarding
the character of the prominent features of the spectrum.
}

\keywords{liquid water, X-ray spectroscopy, multilevel coupled cluster, coupled cluster, water clusters}

\maketitle

\section{Introduction}\label{sec1}

The structure of liquid water has long been a subject
of controversial debate~\cite{guo2002x, wernet2004structure},
and various experimental and theoretical
studies have been conducted to elucidate the matter~\cite{fransson2016x, smith2017soft, jordan2020attosecond}.
X-ray absorption (XA) spectroscopy has emerged as a valuable tool in this respect,
offering insights into the local structure of the molecular environment of
liquid water~\cite{myneni2002spectroscopic, naslund2005x, gavrila2009time, fuchs2008isotope, wernet2004structure, nilsson2010x, smith2017soft}.
However,
the interpretation of the experimental XA spectra relies heavily on accurate theoretical modeling.
Sophisticated theoretical methods are therefore necessary
to improve our understanding of the spectral
properties of this essential substance~\cite{fransson2016x, cisneros2016modeling}.

Simulation of the XA spectrum of liquid water is challenging for several reasons.
The task can generally be divided into two steps: generating a representative set
of molecular geometries, and accurately modeling the core-excited state.
Both steps are associated with their own theoretical challenges.
The first step,
the generation of molecular geometries,
is complicated by the need for accurate \textit{ab initio} molecular dynamics.
For example,
intermolecular interactions
and the delicate nature of  nuclear quantum effects (NQE) must be accounted for.
Due to its performance-to-cost ratio, density functional theory (DFT)
is generally used for this step.
The second step relies on a highly accurate description of the electronic structure
of the core-excited molecule and its close surroundings.
However, achieving this level of accuracy comes at a significant cost,
which typically limits the possibilities of studying bulk properties.
Once again,
DFT has generally been the method of choice.
Overall,
significant differences in the agreement between experimental and theoretical XA spectra
of liquid water have been observed~\cite{prendergast2006x, iannuzzi2008x, chen2010x}.

Due to its attractive computational efficiency 
transition-potential DFT (TP-DFT)~\cite{triguero1998calculations},
has been extensively used to compute XA spectra of liquid water~\cite{leetmaa2010theoretical, fransson2016requirements, iannuzzi2008x}.
However,
choosing an appropriate core-hole potential
(half-core-hole (HCH)~\cite{triguero1998calculations},
full-core-hole (FCH)~\cite{hetenyi2004calculation},
excited-state-core-hole (XCH)~\cite{prendergast2006x})
is non-trivial and depends on the investigated system.\cite{leetmaa2010theoretical, martelli2022properties}
In contrast,
time-dependent DFT (TDDFT) has rarely been used
to simulate XA spectra of liquid water.~\cite{brancato2008accurate, fransson2016requirements}
Generally, TDDFT is known to systematically underestimate
core-excitation energies~\cite{besley2010time,besley2020density, norman2018simulating, rankine2021progress, bussy2021first, besley2012equation}.
However, recently, Carter-Fenk \textit{et al.}~\cite{carter2022electron,carter2022choice}
demonstrated that TDDFT can provide an adequate description of core-excited states by explicitly
accounting for orbital relaxation.
Despite the considerable improvement observed using this electron-affinity TDDFT (EA-TDDFT),
significant discrepancies with experiment still remain~\cite{carter2022electron}.

Another approach that has been used to study the XA spectrum
of liquid water~\cite{vinson2012theoretical, chen2010x, tang2022many}
derives from the Green's function framework
for the Bethe--Salpeter equation (BSE)~\cite{nakanishi1969general}
using the GW approximation~\cite{hedin1965new,hedin1999correlation, reining2018gw}.
This GW-BSE method~\cite{onida2002electronic, leng2016gw, vinson2011bethe}
found its original application in the field of solid state physics.~\cite{blase2018bethe}
Whereas the GW approximation provides a significant correction to the quasi-particle states,
BSE can describe excitonic effects accurately.
In the recent study by Tang \textit{et al.}~\cite{tang2022many},
an excellent match of simulated and experimental spectra are reported.
Their approach includes the use of self-consistent quasi-particle
wave functions and approximate inclusion of the coupling between
the core and high-lying valence excitations.
In this way, increased pre- and main-edge intensities are obtained,
compared to more standard GW-BSE calculations excluding these effects.

Coupled cluster theory offers a highly accurate description of
electronic excitations and spectroscopic properties of molecular systems.
Such accuracy comes at a great expense,
due to the steep polynomial scaling~\cite{helgaker2013molecular}.
The molecular system size is therefore severely limited.
The most widely used model is the coupled cluster singles and doubles (CCSD) approach.
However,
it is often necessary to use a more accurate coupled cluster method
when modeling X-ray spectroscopy.
This is primarily because the excitations from a core orbital result in
strong orbital relaxation,
which can be captured via, e.g., triple excitations in the wave function~\cite{norman2018simulating,coriani2012coupled,CARBONE2019241,eomcc_triples}.

The high computational cost of the coupled cluster models
is the reason for the limited number of such studies
on liquid water~\cite{fransson2016x}.
In a study by Fransson \textit{et al.}~\cite{fransson2016requirements},
the X-ray absorption spectra of small water clusters (up to trimers)
were computed using Lanczos-based coupled cluster
damped linear response theory~\cite{coriani2012coupled,coriani2012lanczos}.
They used coupled cluster the singles and perturbative doubles (CC2)~\citep{christiansen1995second}
and CCSD models.
List \textit{et al.}~\cite{list2014lanczos} also used
a similar response framework together with
polarizable embedding~\cite{peccsd} to compute the XA spectrum of an extended water system.
Here,
only a single water molecule was treated quantum mechanically,
whereas the rest was described through polarizable embedding.
Note that both studies were conducted before the introduction of the core-valence
separation approximation in coupled cluster theory~\cite{CVS_CC,CVS_CC_erratum},
and hence substantially limited in the size of water clusters that could be treated.
Within the core-valence separation
approximation, only core excitations are accounted for
when computing the core-level spectra.
Neglecting valence excitations while computing the core excitations has been
justified by the large energetic and spatial separation between core and valence orbitals~\cite{CVS}.
Utilizing the core-valence separation in coupled cluster calculations
significantly facilitated simulations of core-excited states.

As delocalized intermolecular excitations
are important when describing XA of liquid water,
a study where several water molecules are treated with coupled cluster
could greatly benefit the research field.
Given its high accuracy,
coupled cluster theory can shed light on the
composition of the XA spectrum originating from different hydrogen bond
configurations present in liquid water.
As previously pointed out~\cite{VazdaCruz2019},
one challenge in such analyses
is the strong reliance on transition dipole moments and
transition energies,
whose accuracy is limited by the methodology and the system size.

One strategy to reduce the cost of coupled cluster calculations,
while retaining high accuracy,
is to use a high-level coupled cluster method only in an active region of the molecular system,
while describing the rest at a lower level of coupled cluster theory.
This approach defines the multilevel coupled cluster (MLCC) methods~
\cite{myhre2013extended, myhre2014multi, myhre2016multilevel, myhre2016multilevelXray,folkestad2019multilevel, folkestad2020equation, folkestad2021multilevel, paul2022oscillator},
where the
molecular system is partitioned into active and inactive orbital spaces
treated with different coupled cluster models.
The entire molecular system is described by a single wave function~\cite{myhre2016multilevel}.
The multilevel strategy can be further extended by considering
a subset of the inactive orbitals at the mean-field level.
This approach is called \textit{coupled cluster in Hartree--Fock} (CC-in-HF)~\cite{folkestad2021multilevel,eTprog}.
For the partitioning of the orbital space
there are two options to consider.
One option is to use localized orbitals in the active region.
Alternatively, correlated natural transition orbitals (CNTOs)~\cite{hoyvik2017correlated, folkestad2019multilevel, paul2022oscillator} can be used.
The latter choice generally yields higher accuracy at a lower cost, as the resulting
active orbitals
are tailored to compactly describe excitations of interest. The degree of (de-)localization
of the active orbital space is then determined by the character of the transitions.
Active space approaches,
such as MLCC and CC-in-HF,
are particularly suitable to simulate XA spectra,
because core excitations are localized.
Theoretical carbon, oxygen and nitrogen
\textit{K}-edge spectra computed using these methods
are in excellent agreement with experimental
results~\cite{myhre2016multilevel, myhre2016multilevelXray,folkestad2020equation}.
It has also been demonstrated that these methods may be very useful for applications to large and solvated systems~\cite{folkestad2021multilevel}.

In the present work,
for the first time, multilevel coupled cluster methods
for core excitations are applied to an extensive sampling
of the configuration space
to model the XA spectrum of liquid water.
State-of-the-art
path-integral molecular dynamics (PIMD)~\cite{cp2k,ceriotti2016nuclear, thomsen2021nuclear}
is used to include nuclear quantum effects.
The strongly constrained and appropriately normed (SCAN) DFT functional
is employed in the PIMD simulations~\cite{sun.2015}.
We use the multilevel coupled cluster singles, doubles, and
perturbative triples
in Hartree--Fock (MLCC3-in-HF)~\cite{myhre2016multilevel, paul2022oscillator} model, in addition to the less accurate CCSD in Hartree--Fock (CCSD-in-HF) model.
\begin{figure}[ht]
    \centering
     \includegraphics[width=0.49\linewidth]{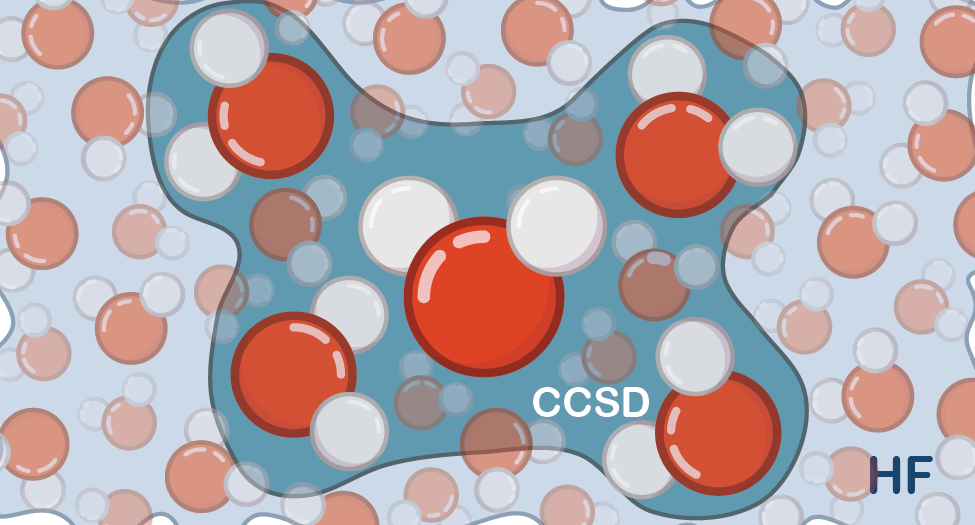}
     \includegraphics[width=0.49\linewidth]{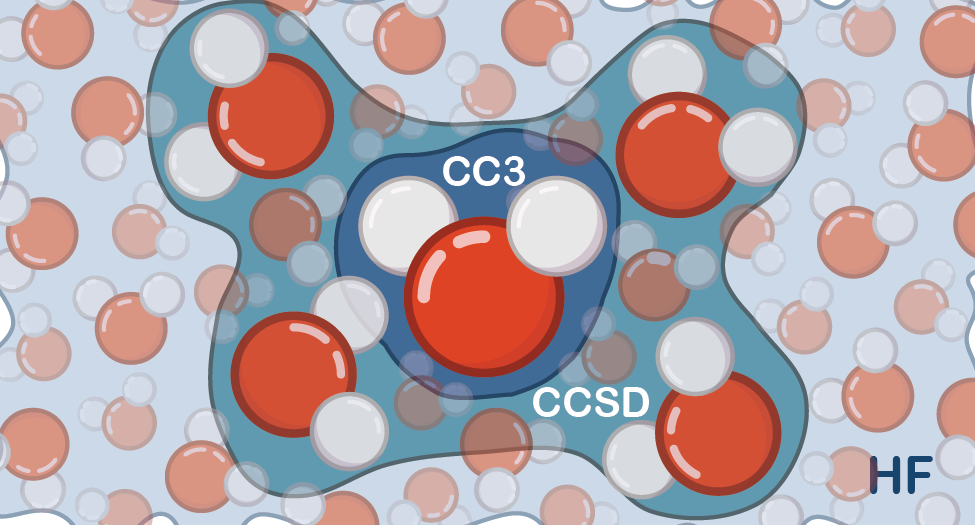}
    \caption{Illustration of the subsystems used in CCSD-in-HF (left) and MLCC3-in-HF (right).
    The region containing the five water molecules is referred to as CCSD space,
    whereas the region around the central water molecule on the right is denoted CC3 space.}
    \label{fig:method_illustration}
\end{figure}
In CCSD-in-HF, CCSD is restricted to orbitals localized on
the five central water molecules of a water cluster.
In the following we refer to this active space as the \textit{CCSD space}.
The remaining orbitals are treated at the Hartree--Fock level;
the partitioning of the system is illustrated in Fig.\,\ref{fig:method_illustration}.
MLCC3-in-HF is built upon CCSD-in-HF, where we choose an additional subset of orbitals
which is treated with CC3;
see the right panel of Fig.\,\ref{fig:method_illustration}.
In the following we refer to this additional active space as \textit{CC3 space}.

\section{Results and Discussion}\label{sec2}

The XA spectrum of liquid water
is usually divided into three regions:
a distinctive pre-edge region centered around $\SI{535}{\eV}$,
a main-edge (also referred to as near-edge) region at $\qtyrange{537}{538}{\eV}$, and
a post-edge at $\qtyrange{540}{542}{\eV}$~\cite{fransson2016x, nilsson2010x,sellberg2014comparison}.
The pre- and main-edges are the predominant features of the spectrum.
The pre-edge peak is attributed
to a highly localized excited state and its intensity is ascribed to
distorted hydrogen bonds~\cite{wernet2004structure, tokushima2008high, fransson2016requirements, cavalleri2002interpretation}
and local liquid disorder~\cite{prendergast2006x, chen2010x, sun2018electron}.
The main- and post-edges are generally attributed to intermolecular delocalized excitations~\cite{brancato2008accurate, tang2022many}.

\begin{figure}
    \centering
    \includegraphics[width=\linewidth]{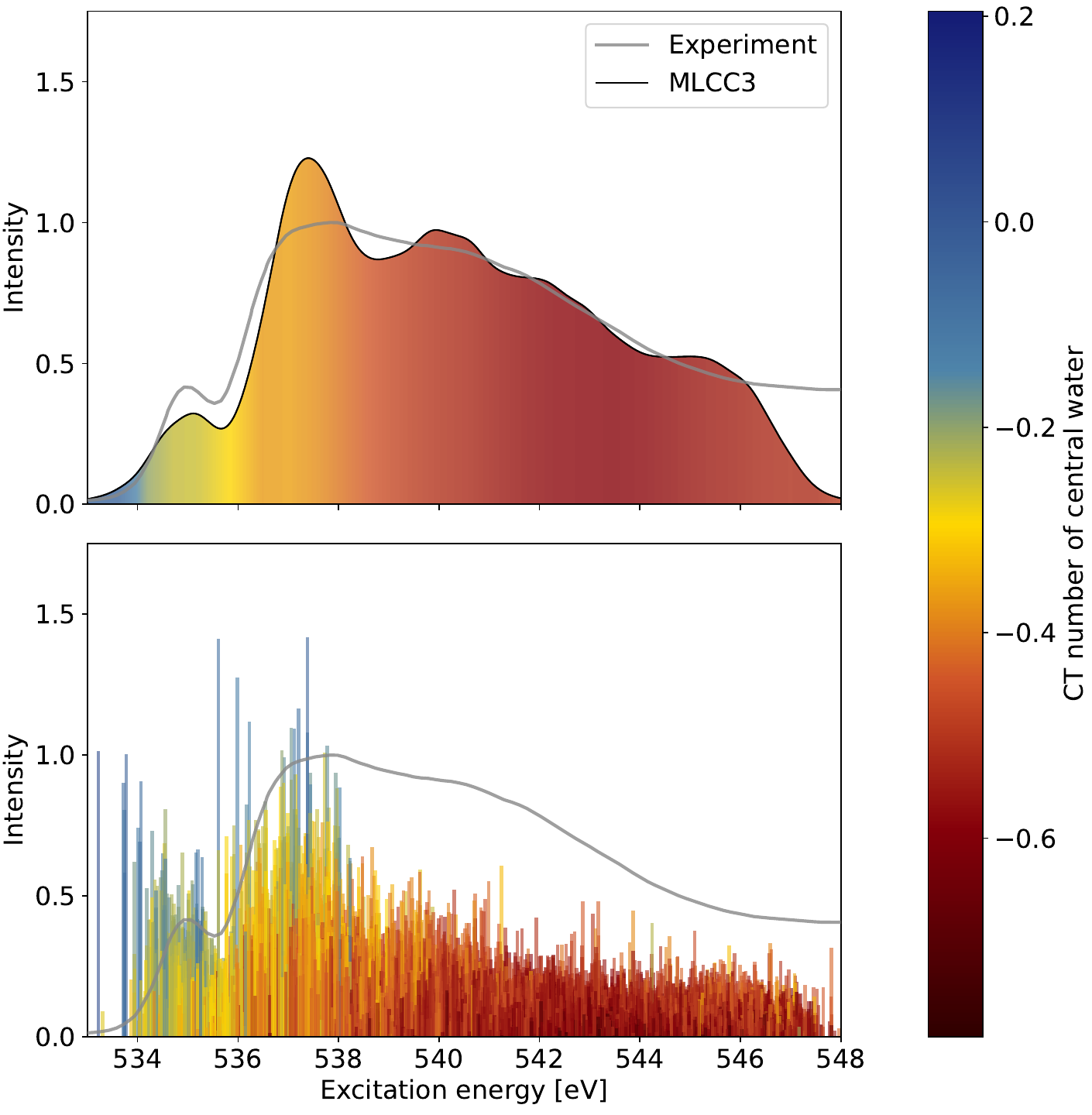}
    \caption{
        Upper panel:
        XA spectrum of water clusters at the MLCC3-in-HF level
        The color gradient represents the charge transfer character,
        related to the influence of the environment.
        Note that the charge transfer (CT) character was only calculated for half the snapshots.
        Lower panel:
        Stick spectrum for the excitations included in the calculation of the charge transfer character.
        The oscillator strengths are scaled by 40 to improve visibility.
        The experimental data was adapted from Ref.~\citenum{experimental_data}.
    }
    \label{fig:mlcc3_density_analysis}
\end{figure}

In Fig.\,\ref{fig:mlcc3_density_analysis},
we present the MLCC3-in-HF XA spectrum of liquid water,
generated from 45 core excited states
for each of the 896 water cluster geometries.
The spectrum is broadened using Voigt profiles with constant FWHM of $\SI{\sim0.6}{\eV}$.
Further details on the molecular dynamics trajectory and the generation of the spectra
can be found in Section \ref{sec3}.

We notice a close agreement of the MLCC3-in-HF results with experiment in both absolute energies and spectral shape,
especially considering the simplistic broadening scheme employed.
The intensity of the pre-edge is slightly underestimated in MLCC3-in-HF.
The main-edge feature is relatively narrow and intense,
but the post-edge shows excellent agreement with the experimental data.
As a consequence, the transition from main- to post-edge appears less smooth than the experiment.
Note that no shift of the energies is needed to align the MLCC3-in-HF spectrum with the experiment.
As shown in Fig.\,\ref{App-fig:mlcc3_convergence},
this is due to cancellation of errors related to the restriction of triple excitations to the CC3 space.
In a full CC3-in-HF calculation,
a shift of $\SI{\sim 0.5}{\eV}$ would be needed to match the experimental pre-edge.
Possible sources of errors that cancel are the finite basis set,
the absence of relativistic effects,
and the limited number of orbitals in the CC3 space.

To evaluate the significance of the immediate environment on the XA spectrum,
the charge transfer character
\cite{ChargeTransferNumbers0,ChargeTransferNumbers1,
ChargeTransferNumbers2,ChargeTransferNumbers3}
for each transition has been analyzed.
The charge transfer number is calculated as the trace
of the difference density
(i.e., the difference between the excited and ground state densities;
for details see \textit{Supplementary Information} Section \ref{SI})
over the atomic orbitals located on the central water molecule.
This partial trace is related to the charge that has been
transferred to, or from, the central water molecule.
Hence, it serves as a measure of the interaction with the surroundings.
A negative value indicates the removal of electrons
while a positive value signifies their addition.
The total trace of the difference density is zero by construction.

Half of the water cluster geometries
were used for the analysis of the charge transfer character.
In Fig.\,\ref{fig:mlcc3_density_analysis},
we assign a color based on the charge transfer number to each excitation.
In the lower panel, we show the individual excitations as bars,
while the upper panel displays the averaged charge transfer character
for all transitions within a bin of size \SI{0.6}{\eV}.
As the excitation energy increases,
there is a distinct transition from low to high charge transfer character.
This is indicated by the gradual change
from positive to increasingly negative values.
This means that in the higher energy region the electron
gradually moves away from the central water molecule.
Because of an artificially reduced density of states (due to the cut-off at 45 excitations),
the gradient becomes slightly brighter around $\SI{546}{\eV}$.
The results of our analysis demonstrate that the excitations of the main-edge
are a mix of relatively localized transitions (blue)
and transitions with some charge transfer character (yellow/orange).
Furthermore, they
are in line with the expectation that the higher
energy excitations will be more diffuse, or delocalized.
Accordingly,
the results show that the environment becomes particularly
important in the post-edge region.
The findings of the charge transfer analysis are corroborated by visualizing
the difference densities for selected states of a single snapshot,
see Fig.\,\ref{App-fig:S68_step01_cc3_difference_34-1_and_4-1} of \textit{Appendix \ref{Appendix}}.

\begin{figure}
    \centering
    \includegraphics[width=\linewidth]{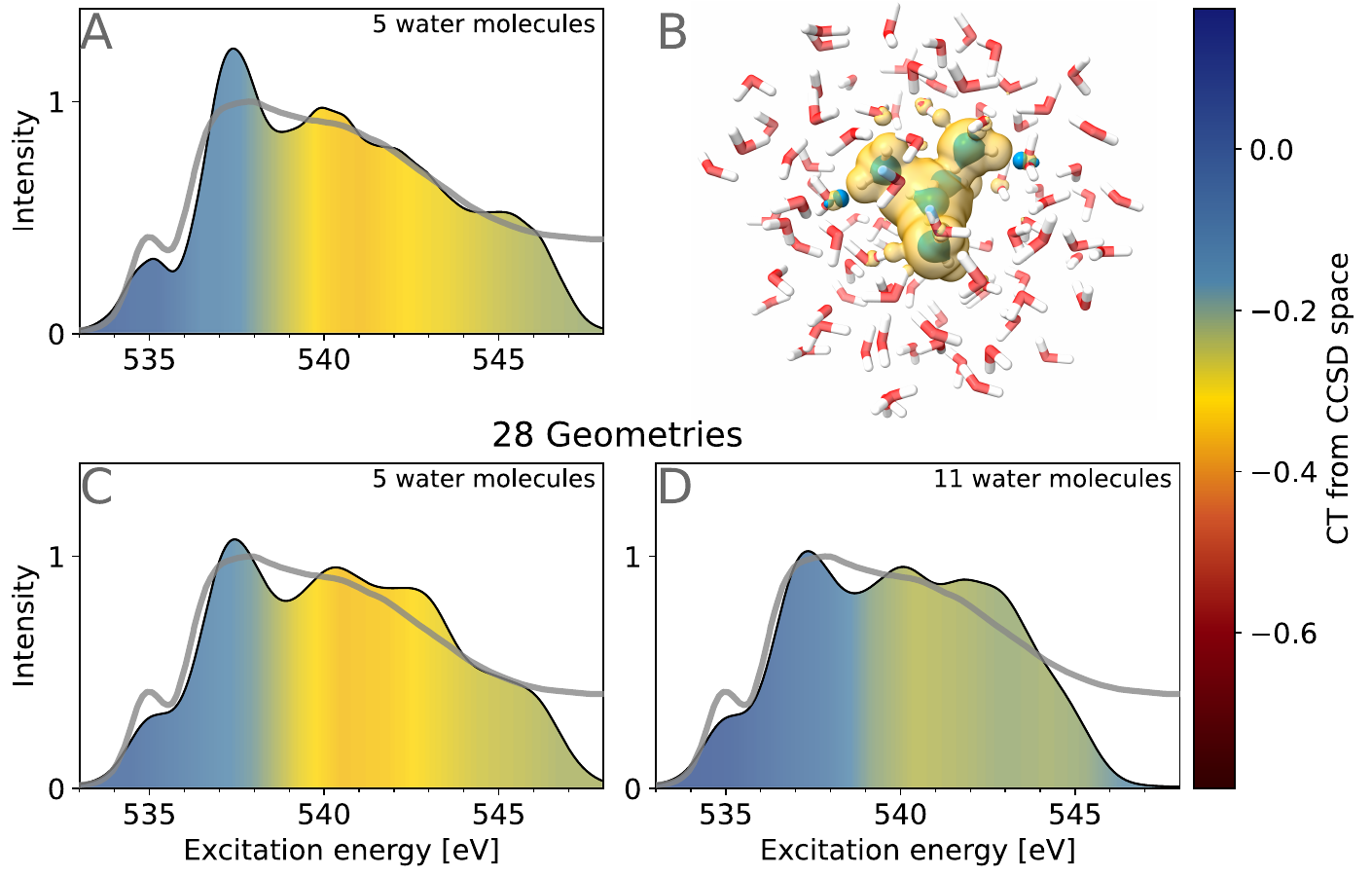}
    \caption{
        Panel A: MLCC3-in-HF charge transfer (CT) character out of the first solvation shell consisting of 5 water molecules.
        Panel B: Occupied (blue) and virtual (yellow) densities
        from the active orbitals used in the CCSD space (iso-value $\SI{0.1}{\isovalue}$).
        Panel C: MLCC3-in-HF charge transfer (CT) character out of the first solvation shell consisting of 5 water molecules
        calculated for 28 selected geometries.
        Panel D: MLCC3-in-HF charge transfer (CT) character out of the first solvation shell consisting of 11 water molecules
        calculated for 28 selected geometries.
        In the panels C and D we broadened the spectrum using Voigt profiles with
        $\SI{0.2}{\eV}$ Lorentzian FWHM and $\SI{0.5}{\eV}$ Gaussian standard deviation.
        The color code here is identical to the one used in Fig.\,\ref{fig:mlcc3_density_analysis}.
        The experimental data was adapted from Ref.~\citenum{experimental_data}.
    }
     \label{fig:density_and_ct}
\end{figure}

Several authors~\cite{list2014lanczos, tang2022many, nilsson2010x} have suggested that
the post-edge is largely characterized by a charge transfer into the second solvation shell and beyond.
As mentioned in the \textit{Introduction},
the coupled cluster calculations presented here are performed in the CCSD space (see Fig. \ref{fig:method_illustration}).
The calculated transitions are thus limited to this active space,
and excitations into the second solvation shell
are enabled only through diffuse components.
To assess how large these components are,
we have plotted the densities from the active occupied and virtual orbitals from the coupled cluster calculations
(panel B of Fig.\,\ref{fig:density_and_ct}).
The water molecules of the second solvation shell
have some virtual density components,
but the density is mainly confined to the first solvation shell.
The quality of the MLCC3-in-HF spectra in relation to the experiment
(Fig.\,\ref{fig:mlcc3_density_analysis})
indicates that such a description is sufficient to
accurately capture
the post-edge feature.
Additionally, we investigated the charge transfer from the first solvation
shell into the other shells,
visualized in panel A of Fig.\,\ref{fig:density_and_ct}.
Our result underpins findings of previous studies
that it is important to go beyond the first solvation shell
to simulate the post-edge region.
However,
we emphasize that the spectrum can be reproduced using a highly accurate electronic structure
method, a diffuse basis set on the first solvation shell,
and a quantum mechanical
embedding for the second solvation shell and beyond.

To further validate the reliability of our results
we performed calculations
with a larger CCSD space, containing 11 water molecules.
Only a small subset of 28 geometries was used and the resulting XA spectrum
is shown in panel D of Fig.\,\ref{fig:density_and_ct}
and Fig. \ref{App-fig:5_vs_11_waters}.
The comparison of the two active spaces (panel C and D)
reveals a small redistribution
of intensity from the main- to the post-edge.
Additionally, the spectrum is contracted,
with the intensity reaching zero at $\SI{546}{\eV}$
due to an increased density of states in the post-edge region.
We further observe a strong reduction in the charge transfer number
when 11 water molecules are included in the CCSD space.
This indicates that most of the interaction with the surrounding water molecules
is now contained in the system.
Overall, the observed differences in the spectral shape are small and do not justify
the drastic increase in computational cost.

%
\begin{figure}
    \centering
    \begin{subfigure}[t]{0.48\textwidth}
        \includegraphics[width=\linewidth]{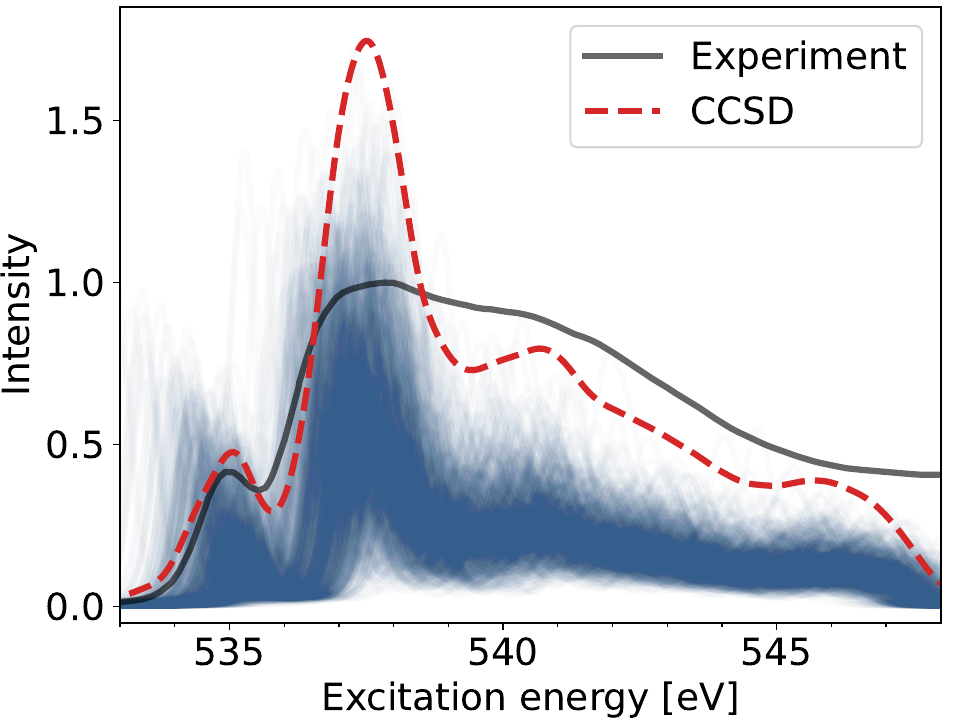}
    \end{subfigure}
    \begin{subfigure}[t]{0.48\textwidth}
        \includegraphics[width=\linewidth]{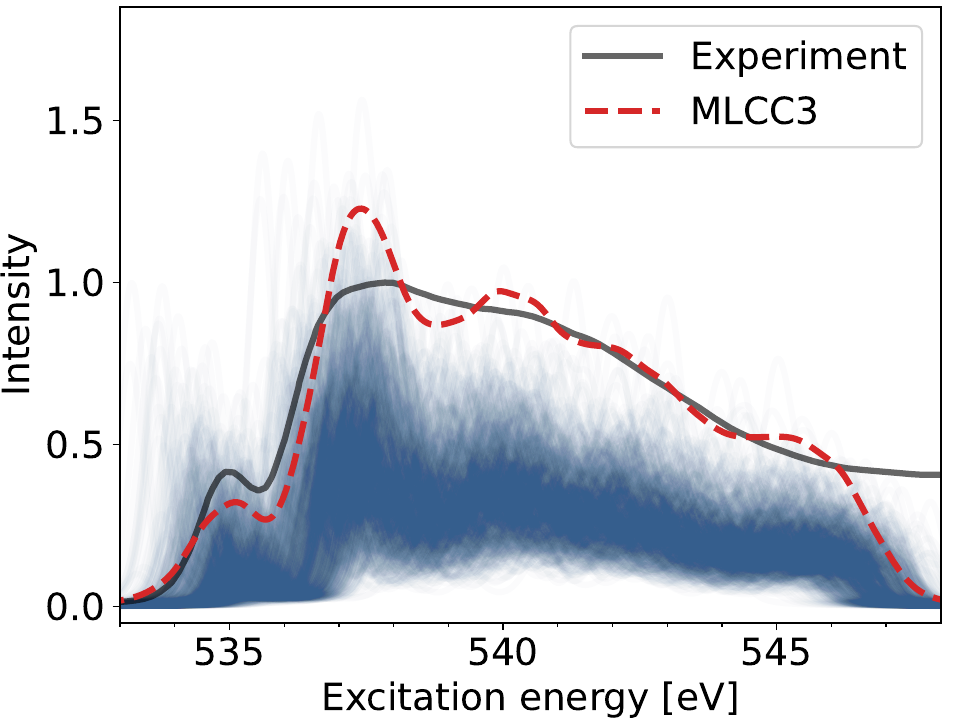}
    \end{subfigure}
    \caption{
        XA spectra of water clusters at the CCSD-in-HF (left) and MLCC3-in-HF (right) level of theory.
        The CCSD-in-HF spectrum was shifted by $\SI{-1.5}{\eV}$ to match the experiment.
        The experimental data was adapted from Ref.~\citenum{experimental_data}.
    }
    \label{fig:snapshots_voigt}
\end{figure}
To demonstrate the importance of the triple excitations~\cite{norman2018simulating}
(included in MLCC3-in-HF) for the quality of the obtained XA spectra of liquid water,
we performed CCSD-in-HF calculations (shown in Fig.\,\ref{fig:snapshots_voigt}).
The CCSD-in-HF spectrum was shifted by $\SI{-1.5}{eV}$ to align with experiment.
Thus, the absolute peak positions are less accurate than in MLCC3-in-HF.
While both models show excellent relative
peak positions,
the most significant effect of using MLCC3-in-HF
is a notable
improvement in the relative intensities in the spectrum for the main- and post-edge features.
Significant changes of
the intensities in CC3 compared to CCSD
have been observed previously~\cite{paul2020new},
where CC3 predicts lower oscillator strengths than CCSD.

From Fig.\,\ref{fig:snapshots_voigt}, we observe that
the pre-edge calculated with CCSD-in-HF is sharper than that of MLCC3-in-HF
and its intensity is slightly
overestimated compared to experiment.
The CCSD-in-HF main-edge is characterized by a narrow peak
twice as intense as any other area of the spectrum.
Relative to the main-edge,
the intensity of the post-edge is underestimated by CCSD-in-HF.
In MLCC3-in-HF this underestimation is corrected
due to the smaller oscillator strengths in the main-edge
and an increased density of states at high energies
compared to CCSD-in-HF.
The increased density of states originates from a stabilization of doubly excited states
due to the inclusion of triple excitations.
Both methods exhibit a peak
in the post-edge, above $\SI{540}{\eV}$ followed by a smooth decrease
in intensity up to $\SI{545}{\eV}$.
The excitations contributing to the post-edge lie
above the ionization limit.
Therefore a special treatment may be required, e.g.,
using continuum functions or Stieltjes imaging \cite{fransson2016requirements}.
However, the quality of the MLCC3-in-HF results indicate that this might not be necessary.

Further analysis of the computed XA spectra reveals that the pre-edge is eliminated
when the first excitation is removed from each of the individual spectra
(see Fig.\,\ref{App-fig:wo_first_peak} in \textit{Appendix \ref{Appendix}}).
This excitation is primarily localized on the central water molecule,
as evidenced by the low degree of charge transfer shown in Fig.\,\ref{fig:mlcc3_density_analysis}.
Sampling the contributions to the XA spectrum in the molecular
frame of each water molecule,
we can decompose the XA intensity into irreducible
representations in the C$_{\rm{2v}}$ point group,
see Fig.\,\ref{App-fig:xyz_contributions} in \textit{Appendix \ref{Appendix}}.
We see that limitations in the description of B$_{\rm{2}}$
transitions are predominantly responsible for the too narrow main-edge.

The main spectral regions of the XA spectrum of water
have been associated with different hydrogen bond configurations
in water~\cite{wernet2004structure,hetenyi2004calculation}.
However,
concerns have been raised regarding the contributions of these configurations~\cite{VazdaCruz2019}.
Vaz da Cruz~\textit{et al.} noted that the accuracy of such an analysis relies heavily on the quality of the calculated
energies and transition dipole moments,
which in turn depends on the chosen electronic structure method.
Given that these quantities are highly accurate in MLCC3-in-HF,
we have analyzed the spectral
composition in terms of various hydrogen bond configurations.
First, to determine their distribution
we employed the criterion suggested in the
\textit{Supporting Information} of Ref.~\cite{wernet2004structure}.
In the water clusters,
75\% of the central water molecules act as double donors,
23\% as single donors,
and 2\% do not participate in any donation.
In particular, this corresponds to the following distribution of the
most common donor(D)--acceptor(A) hydrogen bond configurations:
53\% D2A2, 18\% D2A1, 12\% D1A2 and 10\% D1A1.
In Fig.\,\ref{App-fig:h-bond_analysis}
of \textit{Appendix \ref{Appendix}},
we have plotted the averaged spectra for these configurations.
Overall, our findings show qualitative similarities with the results obtained
by Hetenyi \textit{et al.}~\cite{hetenyi2004calculation}
using the FCH DFT approximation.
Comparing to the prevalent D2A2 configuration, breaking a donor bond (D1A2)
results in a relative increase in the intensity of the pre- and main-edges
compared to the post-edge region. Breaking an acceptor bond (D2A1),
on the other hand,
results in a relative increase in the intensity of the pre- and post-edges
compared to the main-edge region.

\section{Methods}\label{sec3}
\subsubsection*{Molecular dynamics}
The quality of the calculated XA spectra relies significantly on the quality
of the underlying geometries.
Assumption of classical nuclei in molecular dynamics
leads to neglecting nuclear quantum effects (NQEs),
particularly important for the
description of systems containing light atoms~\cite{markland2018nuclear}.
Liquid water is a notable example where it has been shown that NQEs
may influence its structural and dynamical properties~\cite{chen2016ab}.
Path-integral \textit{ab initio} molecular dynamics (PIMD)
incorporates NQEs in static properties,
which can be observed in the qualitative changes
of the radial distribution functions compared to those obtained
from standard classical trajectories~\cite{ceriotti2016nuclear, markland2018nuclear}.
The effects of the NQEs introduced by PIMD on the XA spectra are well illustrated by the
comparison in Fig.\,\ref{App-fig:tddft_beads},
where linear response TDDFT has been applied
to compute spectra on both classical and path-integral trajectories.

The molecular structures used to calculate the XA spectra
were extracted from a PIMD trajectory.
The PIMD simulations were performed using the i-PI code~\cite{ipi} together with CP2K~\cite{cp2k},
where the electronic structure is calculated with DFT,
using the SCAN functional.
These simulations were performed in the canonical ensemble at $T = \SI{300}{\K}$
for a system of 32 water molecules under periodic boundary conditions
(box size $L_{\mathrm{box}} = \SI{9.85}{\AA}$).
The volume was held constant such that $\rho = \SI{1}{\g\per\cm\tothe{3}}$.
The equations of motion were propagated for $\SI{35}{\ps}$
using a time step $\Delta t = \SI{0.5}{\fs}$.
Nuclear quantum effects were included with a thermostated ring polymer contraction scheme using 24 replicas~\cite{Habershon.2012}.

Structural properties
of the resulting model
are extensively discussed in a recent study by Herrero \textit{et al.}~\cite{herrero2022connection}
This model is consistent with the tetrahedral coordination of liquid water and
provides an accurate description of the structural and dynamical properties
within the considered temperature range.\cite{herrero2022connection}.

\subsubsection*{Geometry sampling and preparation}
For a selection of 28 independent snapshots,
water cluster geometries were constructed by centering
a sphere of $\SI{9}{\angstrom}$ radius 
on each of the 32 unique water molecules of the unit cell.
This procedure results in a total of 896 water cluster geometries, each containing approximately 100 molecules.
The geometries provide a 75:23:2 ratio of double-donor,
single-donor, and non-donor hydrogen bond configurations
of the central water molecule.

Calculations at the linear response TDDFT level confirm that our selection
of snapshots extracted from the trajectory of a single bead
is sufficient to converge the XA spectrum.
In Fig.\,\ref{App-fig:tddft_beads},
we compare the TDDFT~\cite{Bussy.2021d3s} XA spectra
for different selections of snapshots.
For a discussion of the sampling, we refer the reader to the \textit{Supplementary Information} (see Section \ref{SI}).

\subsubsection*{Electronic structure methods}

In coupled cluster theory~\citep{helgaker2013molecular}, the wave function is defined as
\begin{equation}
    \ket{\mathrm{CC}} = e^{{T}} \ket{\mathrm{HF}},
    \label{eq:cc_wavefunction}
\end{equation}
where $\ket{\mathrm{HF}}$ is the Hartree--Fock reference determinant.
The cluster operator $T$ generates excitations of the reference determinant and can be ordered according to the excitation levels of the operator,
\begin{equation}
   {T} = {T}_1 + {T}_2 + {T}_3 + \dots,
    \label{eq:cc_T}
\end{equation}
where $T_1$ generates single excitations, $T_2$ generates double excitations, and so on.
Truncation of $T$ defines the hierarchy of standard coupled cluster methods.
For example, in CCSD, ${T}$ is truncated after ${T}_2$.
High excitation orders can be approximated using perturbation theory.
The CC3 model~\citep{koch1997cc3} is a notable example where triple excitations are included perturbatively.
The multilevel coupled cluster wave function is defined as in equation \eqref{eq:cc_wavefunction}.
However, excitations of higher order are restricted to an active orbital space.
In MLCC3~\citep{myhre2016multilevel},
the triple excitations are restricted to an active space.

The CC-in-HF models are quantum mechanical embedding approaches
where the density of the target region is correlated with coupled cluster theory,
whereas the inactive region is described by a frozen Hartree--Fock density.
The orbitals that enter the coupled cluster calculation are chosen by
identifying orbitals localized on the target region.
In the MLCC3-in-HF and CCSD-in-HF calculations in this work,
the active space is defined by the central water molecule
and its four nearest neighbour molecules
(see CCSD space in Fig.\,\ref{fig:method_illustration}).
In MLCC3-in-HF,
we further partition the orbitals of the target region using CNTOs~\cite{hoyvik2017correlated},
only a subset is treated at the CC3 level of theory.
Additional details on the partitioning can be found in the \textit{Supplementary Information} (Section \ref{SI}).
Note that, although the orbital space treated with CC3 will be largely
localized on the central water molecule, it will contain the necessary
components to accurately describe excitations into the first solvation shell.
It is important to emphasize that in both methods the entire system is characterized
by a single wave function.

The coupled cluster models exhibit steep polynomial scaling.
The CCSD model scales as $\mathcal{O}(N^6)$,
where $N$ relates to the size of the molecular system,
and CC3 scales as $\mathcal{O}(N^7)$.
The increased scaling,
moving from CCSD to CC3,
is often limiting, and systems that can be treated routinely with CCSD
cannot be treated with CC3.
This is the motivation for introducing the MLCC3 approach.
CC3 quality can be obtained at a cost approaching that of CCSD,
when the active orbital space is appropriate.
The power of such an approach is clear from Fig.\,\ref{App-fig:mlcc3_convergence}.
Even with a small active space, the spectral shape obtained in an MLCC3 calculation mirrors that of the full CC3 calculation,
while differing significantly from the CCSD spectrum.

For each water cluster,
we calculated 45 CCSD-in-HF and MLCC3-in-HF core excited states of the oxygen atom
on the central water using core-valence separation~\cite{CVS_CC, CVS_CC_erratum, CVS}.
This water molecule,
and its four closest neighbors,
were treated at the coupled cluster level of theory.
The remaining water molecules entered the calculation through a frozen Hartree--Fock density.
For the central water molecule,
we used the aug-cc-pVTZ basis,
while aug-cc-pVDZ was used for the other four water molecules in the CCSD space.
For the water molecules in the Hartree--Fock space,
we adopted the cc-pVDZ basis set~\cite{dunning1989}.
The frozen core approximation is applied to all core orbitals except for the oxygen core orbital on the central water.
The CC3 space consists of 12 occupied and 72 virtual CNTOs.
This active space is sufficient to obtain CC3 accuracy,
as can be seen from Fig.\,\ref{App-fig:mlcc3_convergence}.
The calculations were performed using a development
version of the $e^T$ program~\cite{eTprog}.

\subsubsection*{Broadening and normalization of theoretical spectra}
The theoretical stick spectra are broadened using Voigt profiles
with Lorentzian FWHM and Gaussian standard deviation
of $\SI{0.2}{\eV}$.
This results in a FWHM of $\SI{\sim 0.59}{\eV}$.
The average spectrum is normalized such that the area underneath
the curve matches that of the experiment in the range of \qtyrange{533}{545}{\eV}.
The experimental data shown in the figures was digitized from Ref.~\citenum{experimental_data}
using Web Plot Digitizer \cite{WebPlotDigitizer}.
For comparison, Fig.\,\ref{App-fig:snapshots_lorentzian} and
Fig.\,\ref{App-fig:snapshots_gauss}
show the spectra broadened using Lorentzian functions with FWHM of $\SI{0.2}{\eV}$
and Gaussian functions with FWHM of $\SI{0.6}{\eV}$.
Despite changing the broadening scheme,
the spectral shape,
i.e. the main features of the spectrum,
remain unchanged.

\section{Conclusion}\label{sec4}
In this work, we have presented the first
large-scale application of the multilevel coupled cluster method, MLCC3-in-HF.
Because of the
high computational cost,
standard coupled cluster models have limited applicability for solvated systems,
except in combination with embedding approaches.
Typically, classical embedding,
such as a molecular mechanics or a polarizable continuum embedding,
is used.
However, the intricate nature of intermolecular interactions of liquid water
is responsible for its unique XA spectrum and, therefore,
an accurate quantum mechanical description is called for.
Our study demonstrates the first successful application of
coupled cluster models to the simulation of XA spectroscopy in liquid water.
The results have broad implications: enabling
accurate simulations of X-ray spectra of solvated chromophores
and investigations of the influence of chemical interactions
in condensed phase using coupled cluster methods.

Our results indicate that
including triple excitations in the coupled cluster wave function,
as in MLCC3-in-HF,
is paramount to correctly reproduce
the experimental intensity ratio between the main- and post-edge.
In contrast to CCSD-in-HF, which contains only double excitations,
MLCC3-in-HF shows a redistribution of intensities originating
from smaller oscillator strengths
and an increased density of states in the
post-edge region.
Due to the inclusion of triple excitations we are also able
to reproduce absolute peak positions with unprecedented accuracy.

While the main-edge of the XA spectrum of liquid water
has generally been attributed to excitations into the first solvation shell,
several authors have suggested that the inclusion of
excitations into the second solvation shell or beyond is crucial
for the accurate reproduction of the post-edge.
We perform the analysis of the charge transfer character
for the excitations in the XA spectrum and confirm
that the excitations of the main-edge are delocalized over the first solvation shell,
whereas the post-edge has contributions beyond it.
However,
we demonstrate that the effective modeling of the
post-edge is possible even without
explicitly considering
the second
or subsequent solvation shells with higher-order electronic structure theory.
Our findings reveal that the post-edge can be
accurately modeled by a high-level
theoretical description, such as MLCC3, of the first solvation
shell, while describing
the more distant environment using a frozen Hartree--Fock density.
One strategy to further improve the accuracy of the computed XA spectrum
is to expand the coupled cluster space to match the reported
experimental coordination number of 4.7 water molecules
\cite{Pathirannahalage2021,HeadGordon2006,ErrHeadGordon2006}. Calculations on a subset of the cluster geometries,
including 11 water molecules in the coupled cluster region,
show only a slight reduction in the intensity of the main-edge, and  does not justify the significant increase in computational cost.

\backmatter

\bmhead{Supplementary information}

Details of the charge-transfer analysis, orbital space partitioning
and average computational cost
are detailed in Section \ref{SI}.

\bmhead{Acknowledgments}

This work has received funding from the European Research Council (ERC)
under the European Union’s Horizon 2020 Research and Innovation Program
(grant agreement No. 101020016).
S.D.F., A.C.P, and H.K. acknowledge funding from the
Research Council of Norway through FRINATEK (project No. 275506).
{M.O. acknowledges funding from the Swedish Research Council (grant agreement No. 2021-04521), and from the European Union's Horizon 2020 research and innovation programme under the Marie Sk{\l}odowska-Curie (grant agreement No. 860553).
S.C. acknowledges support from the Independent Research Fund Denmark--Natural Sciences, DFF-RP2 (grant No. 7014-00258B).}
We acknowledge computing resources through UNINETT Sigma2
--the National Infrastructure for High Performance Computing
and Data Storage in Norway (project No. NN2962k),
from DeIC -- Danish Infrastructure Cooperation (grant No. DeiC-DTU-N3-2023027), and from the Swiss National Supercomputing Centre (project ID uzh1).

\section*{Data availability}

The water cluster geometries, and the data generated and analyzed in this study are available from Ref.~\citenum{geometries}.

\begin{appendices}

\section{Extended Data}\label{Appendix}
\begin{figure}
    \centering
    \includegraphics[width=0.85\linewidth]{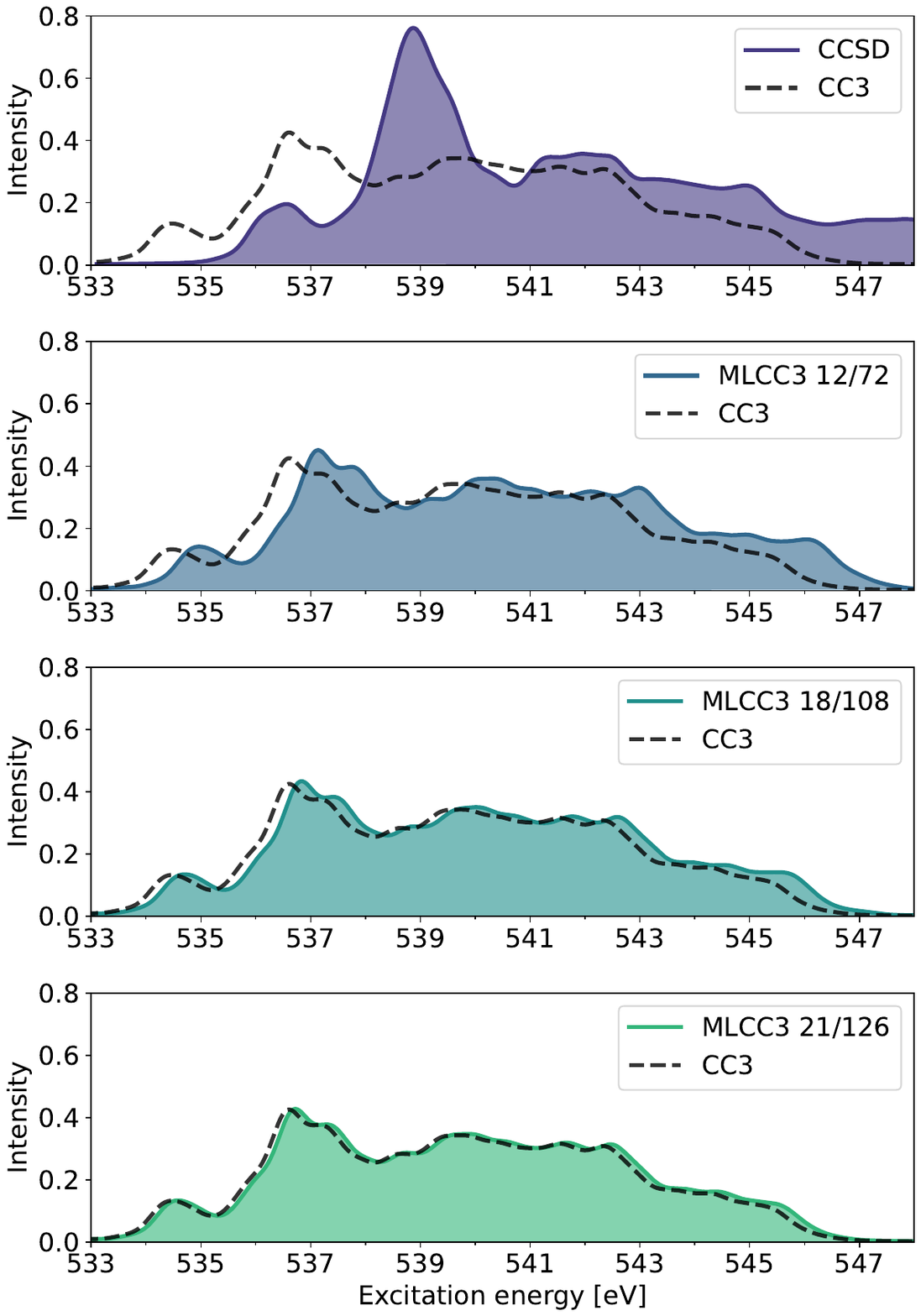}
    \caption{
        Averaged XAS spectra of 28 different water clusters at the MLCC3-in-HF level of theory showing the effect
        of including more CNTOs in the MLCC3 orbital space.
        The MLCC3-in-HF excitations are broadened using Voigt profiles
        with $\SI{0.2}{eV}$ Lorentzian width and $\SI{0.2}{eV}$ Gaussian standard deviation ($\SI{\sim0.59}{eV}$ fwhm).
        The dashed line shows the spectrum obtained for a CC3-in-HF calculation,
        i.e. including all orbitals of the central 5 water molecules in the CC3 calculation.
    }
    \label{App-fig:mlcc3_convergence}
\end{figure}
\begin{figure}
    \centering
    \begin{subfigure}[b]{0.49\textwidth}
        \includegraphics[width=\linewidth]{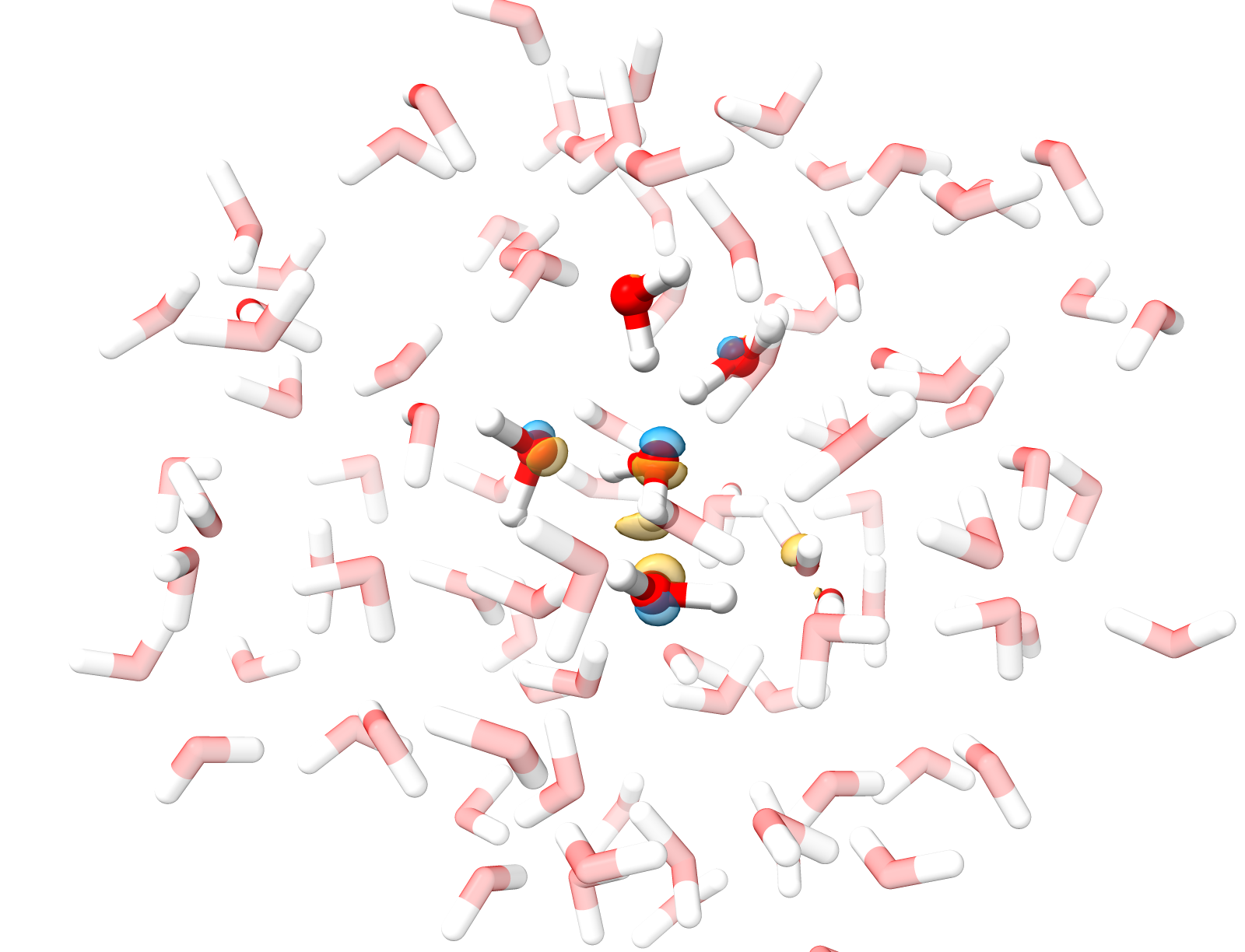}
    \end{subfigure}
    \begin{subfigure}[b]{0.49\textwidth}
        \includegraphics[width=\linewidth]{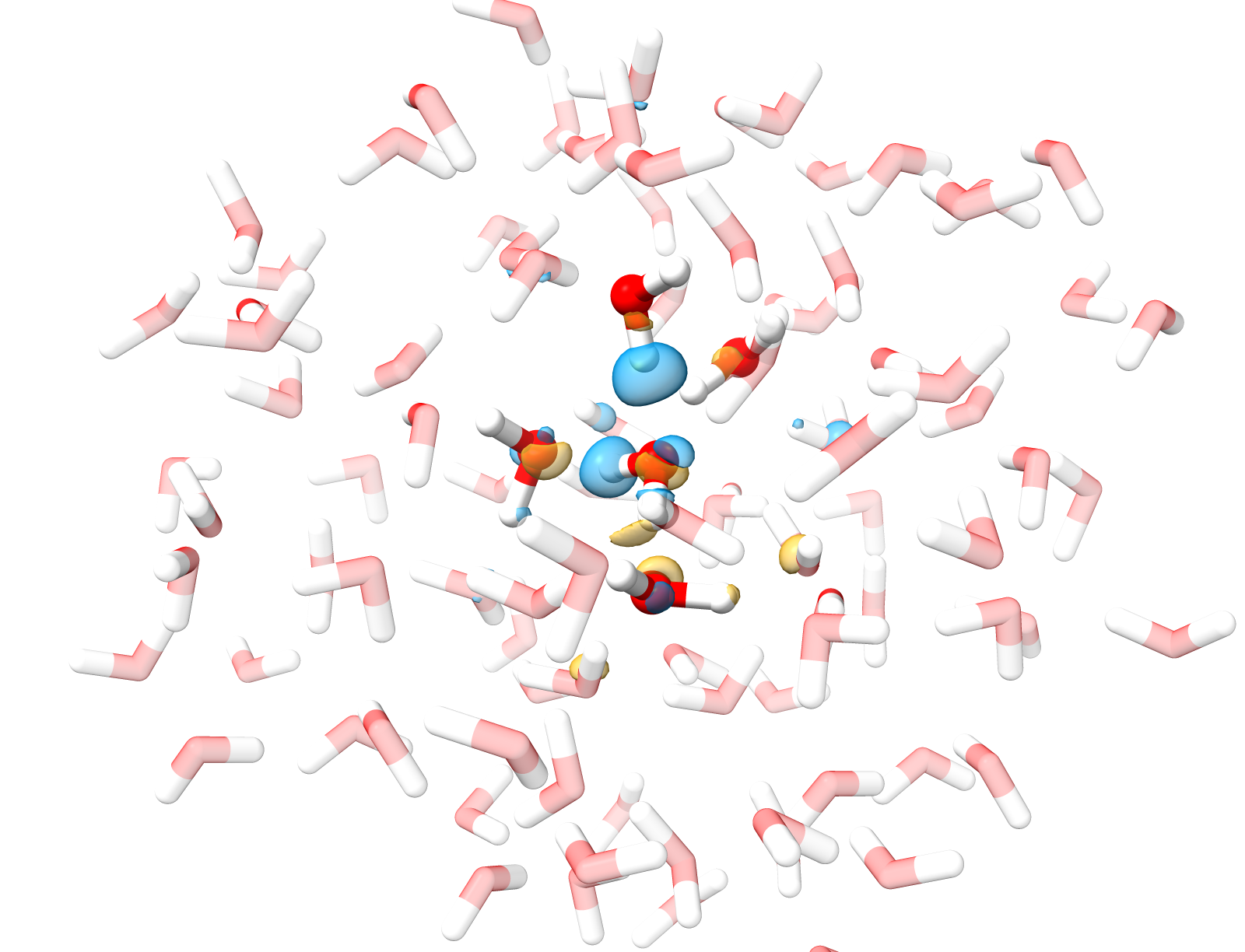}
    \end{subfigure}
    \caption{
        Left: Difference of the MLCC3-in-HF excited state densities of excited states 4 and 1.
        Right: Difference of the MLCC3-in-HF excited state densities of excited states 34 and 1.
        The geometry of snapshot 68 step 01 was used and the
        iso-value was set to $\SI{4e-3}{\isovalue}$.
        The yellow areas indicate a depletion of electron density in the transition from the first excited state (state 1)
        to the final state (either state 4 or state 34), and the blue areas indicate an addition of electron density.
    }
    \label{App-fig:S68_step01_cc3_difference_34-1_and_4-1}
\end{figure}
\begin{figure}
    \centering
    \includegraphics[width=0.7\linewidth]{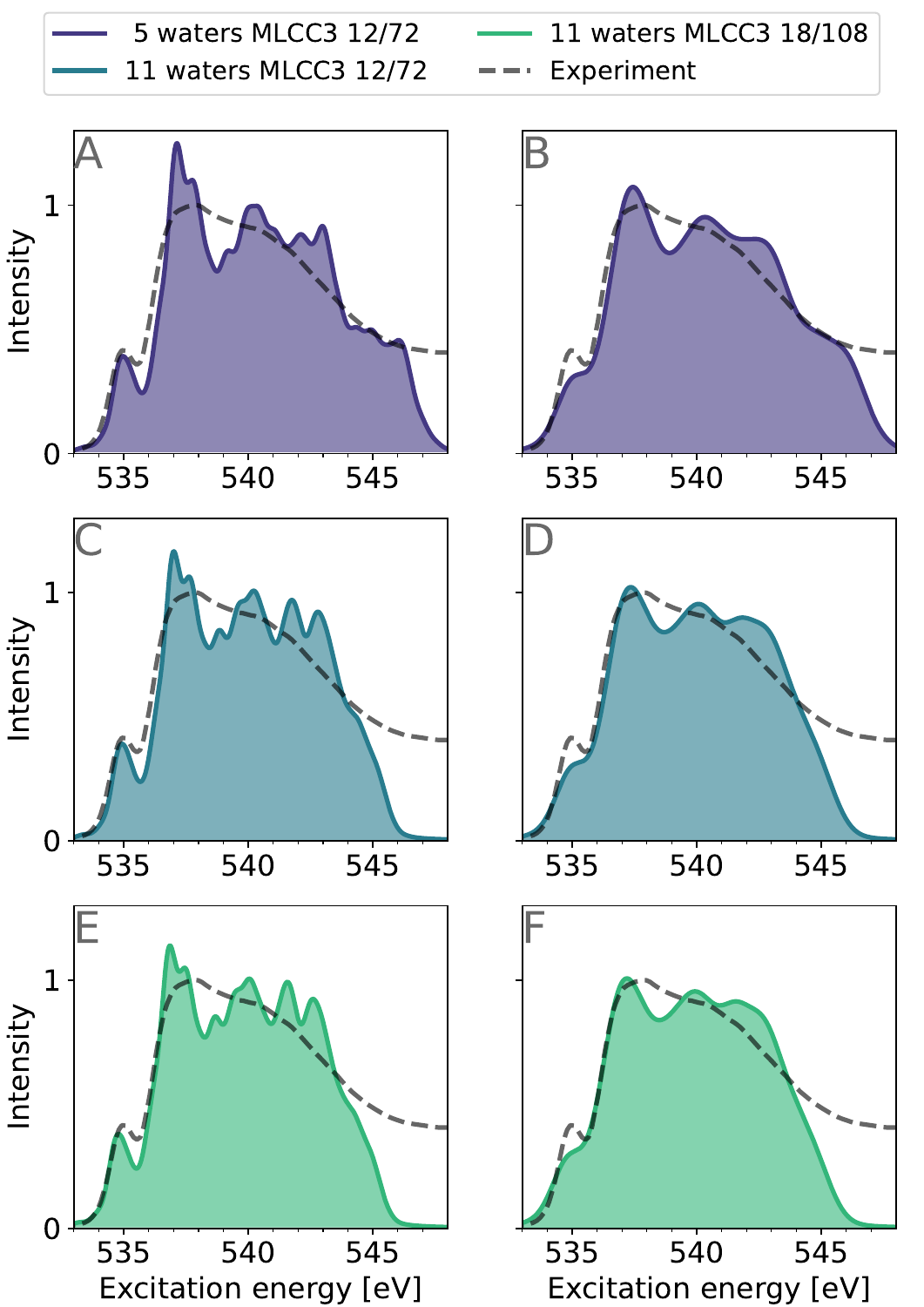}
    \caption{
        XA spectra averaged over 28 different water clusters at the MLCC3-in-HF level of theory showing the effect
        of including more waters in the coupled cluster space and increasing the number of orbitals in the MLCC3 space.
        For the first row (panel A, B) the CCSD space includes the central water
        and the four closest neighbouring water molecules and the MLCC3 space contains 12 occupied and 72 virtual CNTOs.
        For the second row (panel C, D) the 10 closest neighbours were included keeping the number of CNTOs.
        For the third row (panel E, F) the 10 closest neighbours were included and the number of CNTOs was increased
        to 18 occupied and 108 virtual CNTOs.
        In the left column the default broadening was used.
        In the right column the Gaussian standard deviation in the Voigt profiles
        was set to $\SI{0.5}{eV}$ increasing the overall FWHM to $\SI{\sim1.29}{eV}$ to further smooth the spectra.
    }
    \label{App-fig:5_vs_11_waters}
\end{figure}
\begin{figure}
    \centering
    \begin{subfigure}[b]{0.49\textwidth}
        \includegraphics[width=\linewidth]{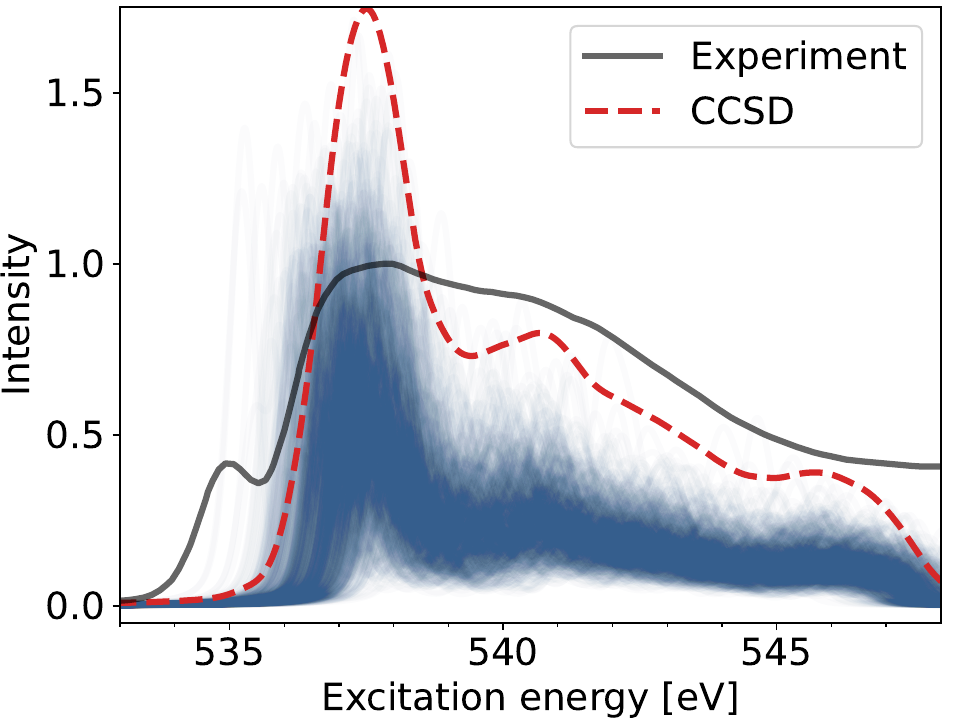}
    \end{subfigure}
    \begin{subfigure}[b]{0.49\textwidth}
        \includegraphics[width=\linewidth]{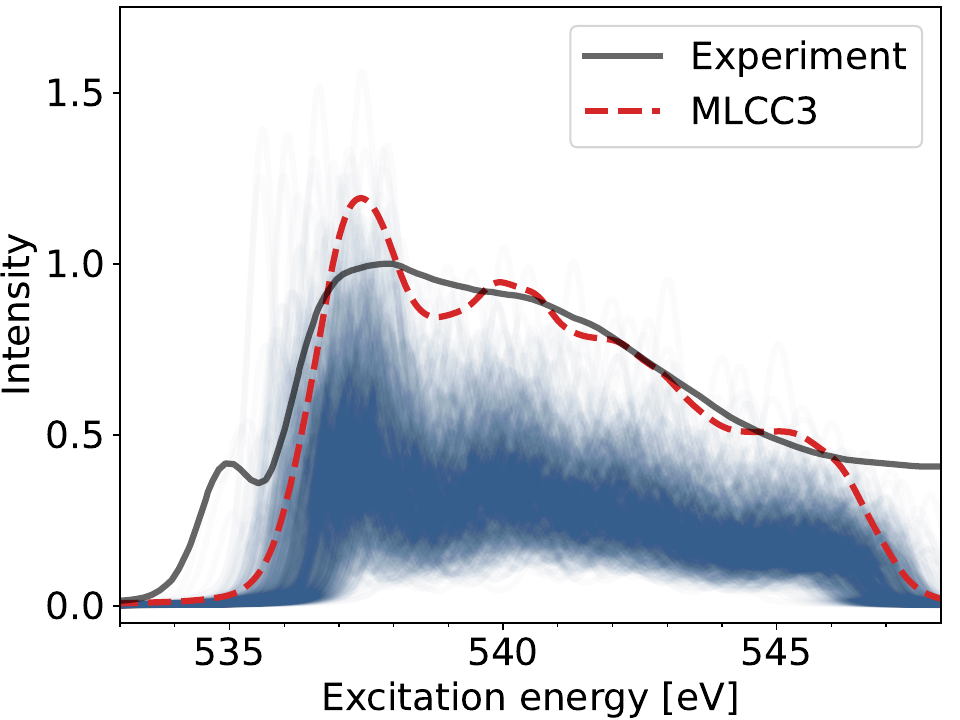}
    \end{subfigure}
    \caption{
        XAS spectra of water clusters at the CCSD-in-HF (left) and MLCC3-in-HF (right) level of  theory.
        The orange lines show the spectra averaged over 896 individual snapshots (blue)
        \textit{The first excitation of each snapshot has been disregarded in this plot to identify the character of
        the pre-edge}.
        The CCSD-in-HF spectrum was shifted by $\SI{-1.5}{\eV}$ to match the experiment.
        The experimental data was adapted from Ref.~\citenum{experimental_data}.
    }
    \label{App-fig:wo_first_peak}
\end{figure}
\begin{figure}
    \centering
    \begin{subfigure}[b]{0.49\textwidth}
    \includegraphics[width=\linewidth]{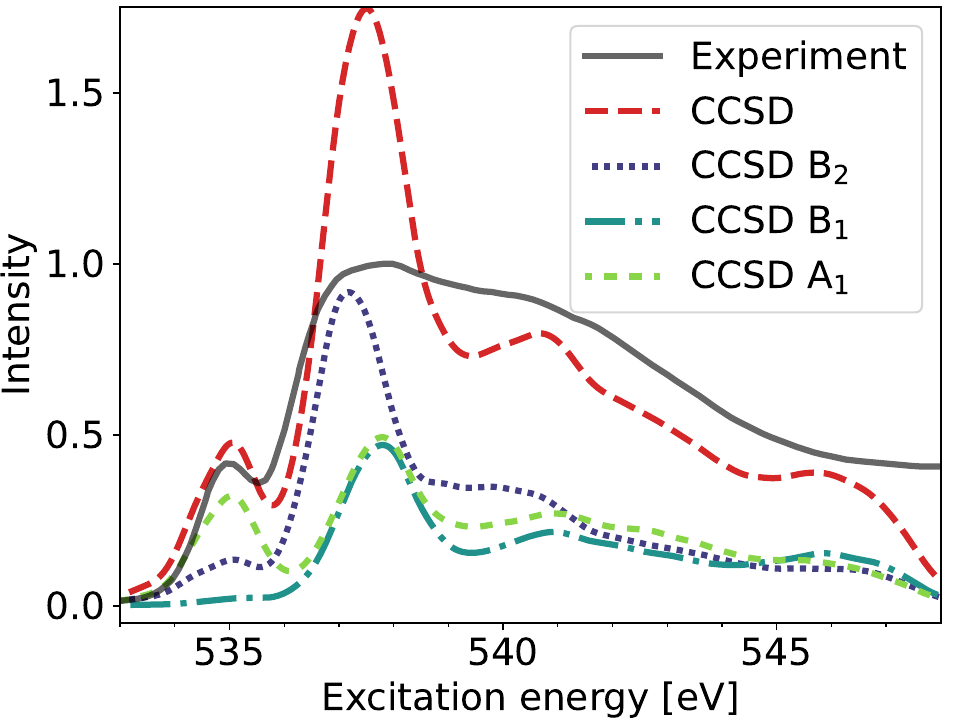}
    \end{subfigure}
    \begin{subfigure}[b]{0.49\textwidth}
        \includegraphics[width=\linewidth]{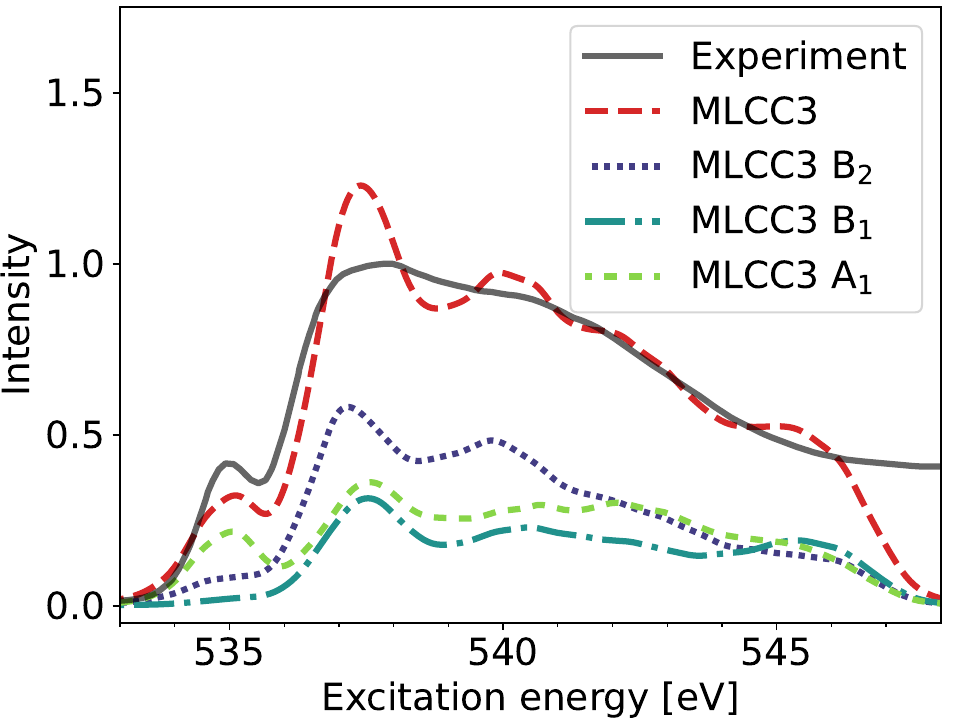}
    \end{subfigure}
    \caption{
        XAS spectra of water clusters at the CCSD-in-HF (left)
        and MLCC3-in-HF (right) level of  theory.
        The orange lines show the complete spectra and the blue,
        red and green lines
        indicate the contributions of excitations that can be related to
        B2, B1 (out of plane) and A1 symmetry in a single water molecule.
        The CCSD-in-HF spectrum was shifted by $\SI{-1.5}{\eV}$ to match the experiment.
    }
    \label{App-fig:xyz_contributions}
\end{figure}
\begin{figure*}
    \centering
    \includegraphics[width=\textwidth]{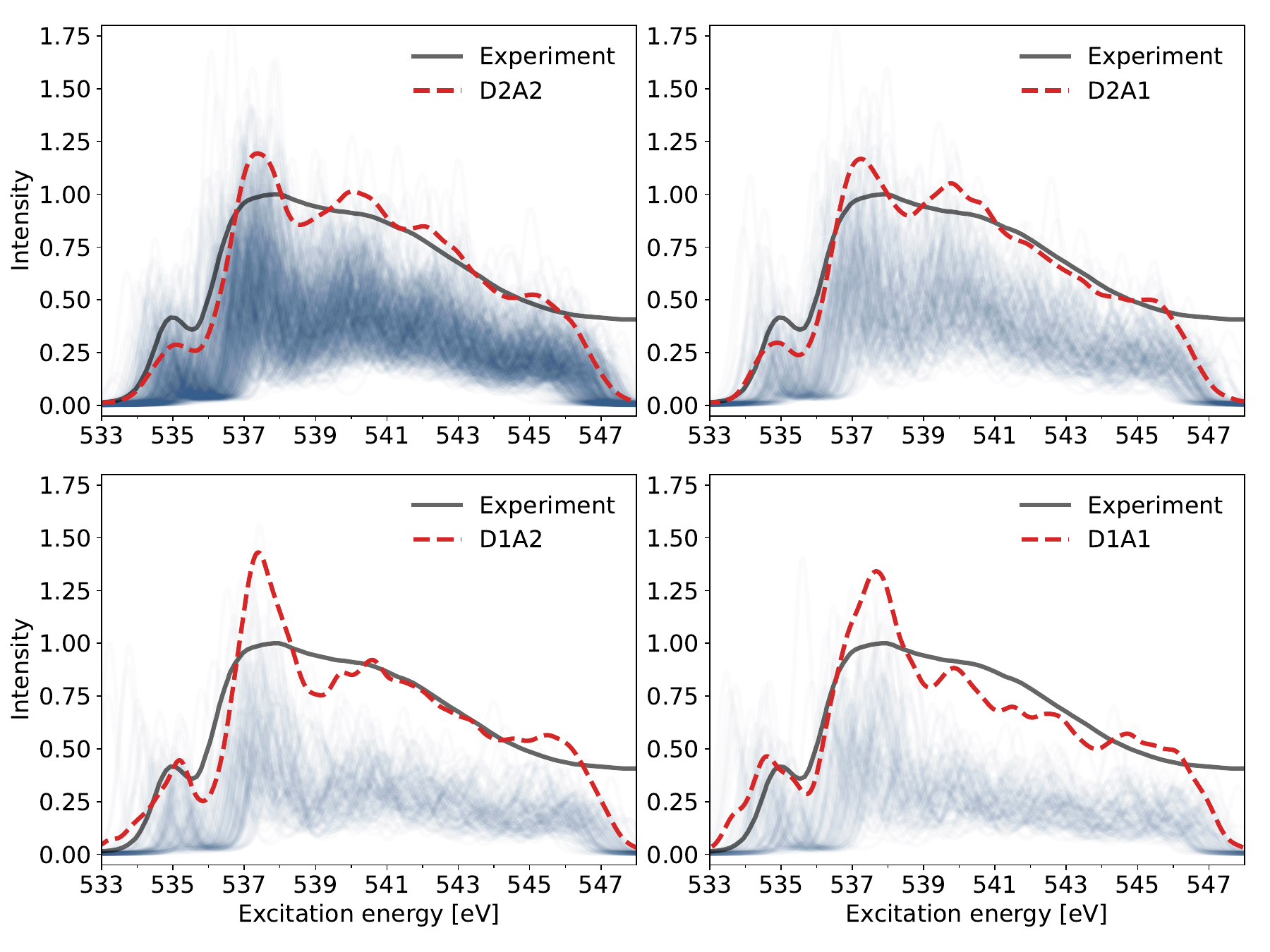}
    \includegraphics[width=0.7\textwidth]{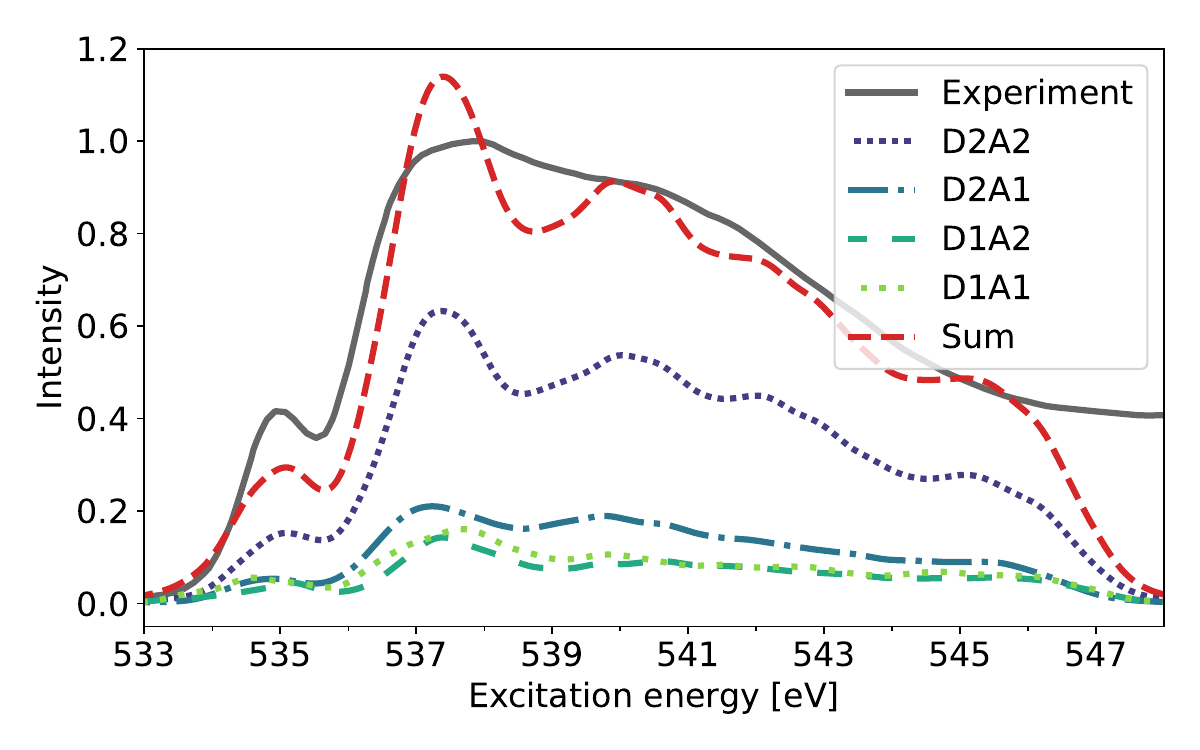}
    \caption{
        XA spectra of water clusters at the MLCC3-in-HF level of  theory,
        divided by hydrogen bond character of the central water(D2A2, D2A1, D1A2, and D1A1).
        The calculated spectra were normalized to the experimental spectrum by
        matching the area under the curves in the range of $\SI{533}{}-\SI{545}{\eV}$
        and a Voigt broadening with Lorentzian FWHM of $\SI{0.2}{\eV}$ and Gaussian standard deviation of $\SI{0.2}{\eV}$ was used.
        The experimental data was adapted from Ref.~\citenum{experimental_data}.
        \textbf{Top:} Individual and averaged spectra.
        \textbf{Bottom:} Spectra scaled according to their importance D2A2: 0.53; D2A1: 0.18; D1A2: 0.12; D1A1: 0.10)
    }
    \label{App-fig:h-bond_analysis}
\end{figure*}
\begin{figure}
    \centering
    \includegraphics[width=0.8\linewidth]{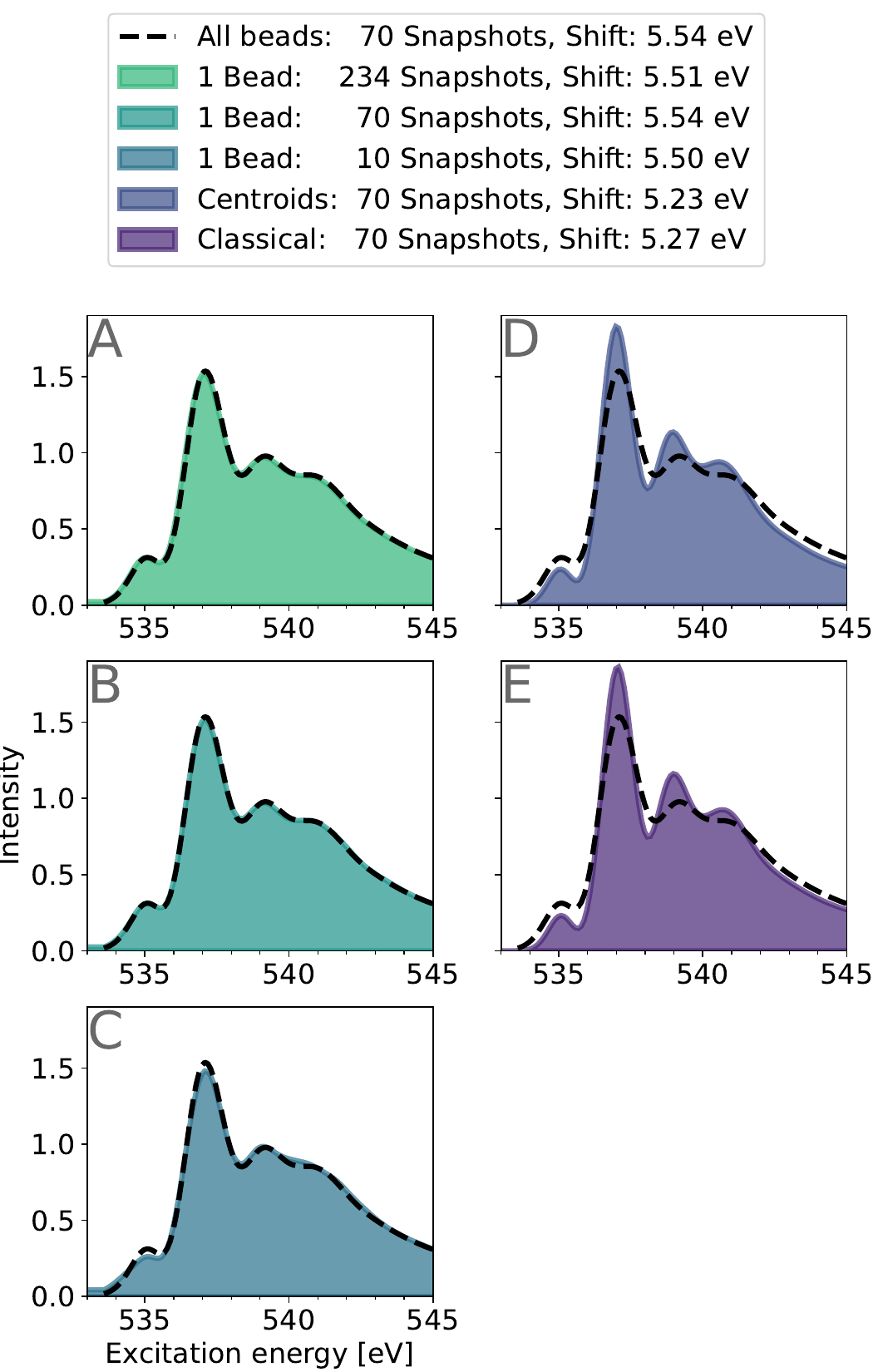}
    \caption{
        Comparison of the XAS spectra computed with LR-TDDFT using different selections
        of snapshots from the trajectory described in the main paper
        and another AIMD trajectory.
        The spectra were broadened using Gaussians of \SI{0.8}{eV} width
        and normalized matching the area of the experimental spectrum in the range $\qtyrange{534}{544}{\eV}$.
        The spectra are averaged over the snapshots and the 32 oxygen atoms in the simulation box.
    }
    \label{App-fig:tddft_beads}
\end{figure}
To evaluate the reliability of the geometries used in our calculations
we investigate different choices for the sampling of the PIMD trajectory
and compare with an AIMD trajectory obtained using the same parameters.
The spectra are computed with LR-TDDFT in the Tamm-Dancoff approximation,
at the PBEh level of theory with the pcseg2 basis set using CP2K~\cite{cp2k}
with periodic boundary conditions~\cite{Bussy.2021d3s}.
The results are shown in Fig.\,\ref{App-fig:tddft_beads}.

First,
we compare the spectra calculated for 70 geometries of the AIMD trajectory (denoted \textit{classical}, panel E)
with the spectrum obtained for 70 geometries from the PIMD trajectory denoted \textit{centroids} (panel D).
For the centroid geometries the hydrogen atoms are placed at the center of the 24-bead ring polymers.
From the lower two panels of Fig.\,\ref{App-fig:tddft_beads},
we see that the spectra are virtually identical
and only differ slightly in the intensity of the post-edge.
However,
both spectra exhibit significant differences compared
to the spectrum obtained using the geometries
of all beads for 70 snapshots (denoted \textit{All beads}).
By including quantum effects for the hydrogen atoms the spectra
become smoother and the pre-edge intensity increases.

Considering all beads amounts to a large number of geometries.
Therefore,
we investigate a sampling of a single bead of the ring polymer instead of all beads.
Calculating the spectrum for a single bead for 10 snapshots (panel C),
already mostly reproduces the spectrum obtained when all beads are taken into account.
When the number of snapshots is increased to 234 (panel A) or even only 70 (panel B),
the difference becomes negligible.
Therefore,
we conclude that considering a number between 10 and 70 snapshots
for a single bead should be sufficient to produce an accurate spectrum.
In our calculations we considered 28 snapshots
for which we calculate 896 individual spectra by sampling core excitations on each oxygen atom.
\begin{figure}
    \centering
    \begin{subfigure}[b]{0.49\textwidth}
        \includegraphics[width=\linewidth]{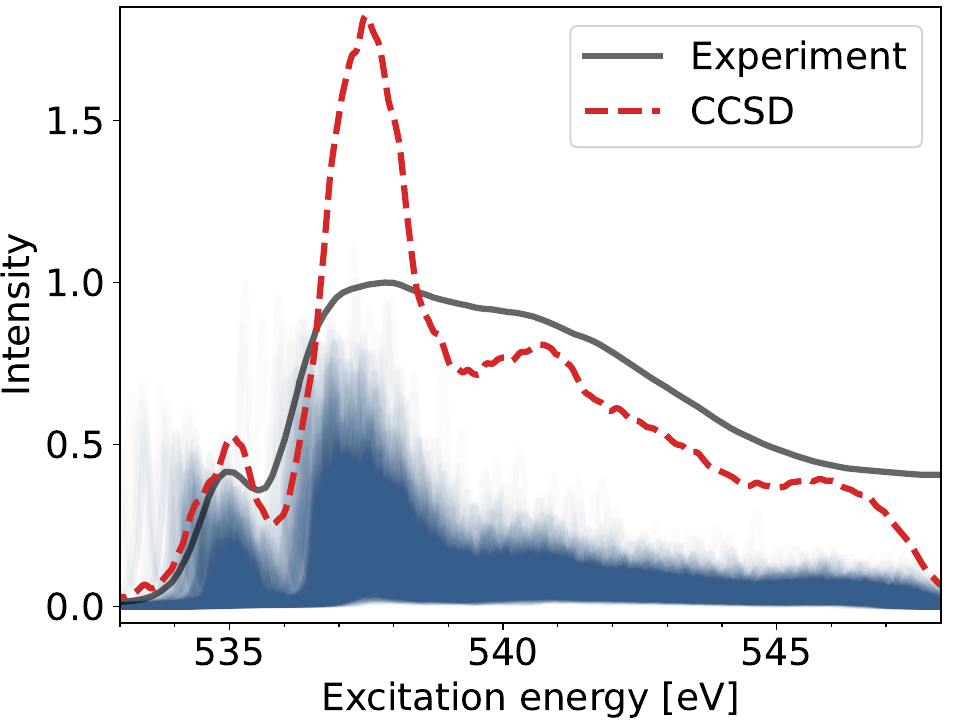}
    \end{subfigure}
    \begin{subfigure}[b]{0.49\textwidth}
        \includegraphics[width=\linewidth]{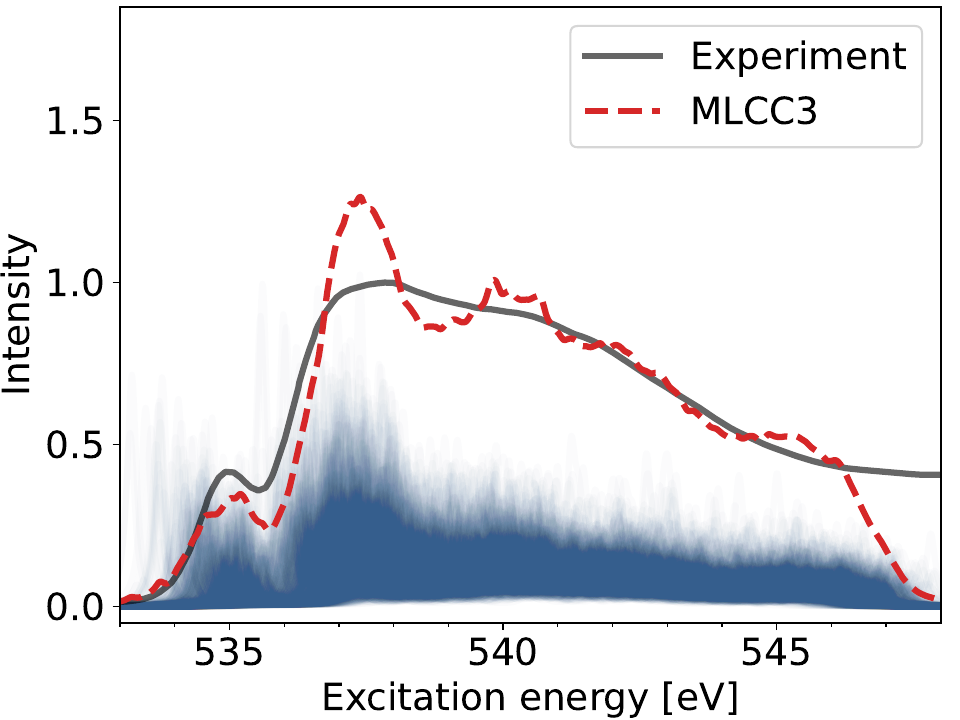}
    \end{subfigure}
    \caption{
        XAS spectra of water clusters at the CCSD-in-HF (left) and MLCC3-in-HF (right) level of  theory.
        The orange lines show the spectra averaged over 896 individual snapshots (blue)
        The calculated spectra were normalized to the experimental spectrum by
        matching the area under the curves in the range of $\SI{533}{}-\SI{545}{eV}$
        and a \textit{Lorentzian broadening} with FWHM of $\SI{0.2}{\eV}$ was used.
        The CCSD-in-HF spectrum was shifted by $\SI{-1.5}{\eV}$ to match the experiment.
    }
    \label{App-fig:snapshots_lorentzian}
\end{figure}
\begin{figure}
    \centering
    \begin{subfigure}[b]{0.49\textwidth}
        \includegraphics[width=\linewidth]{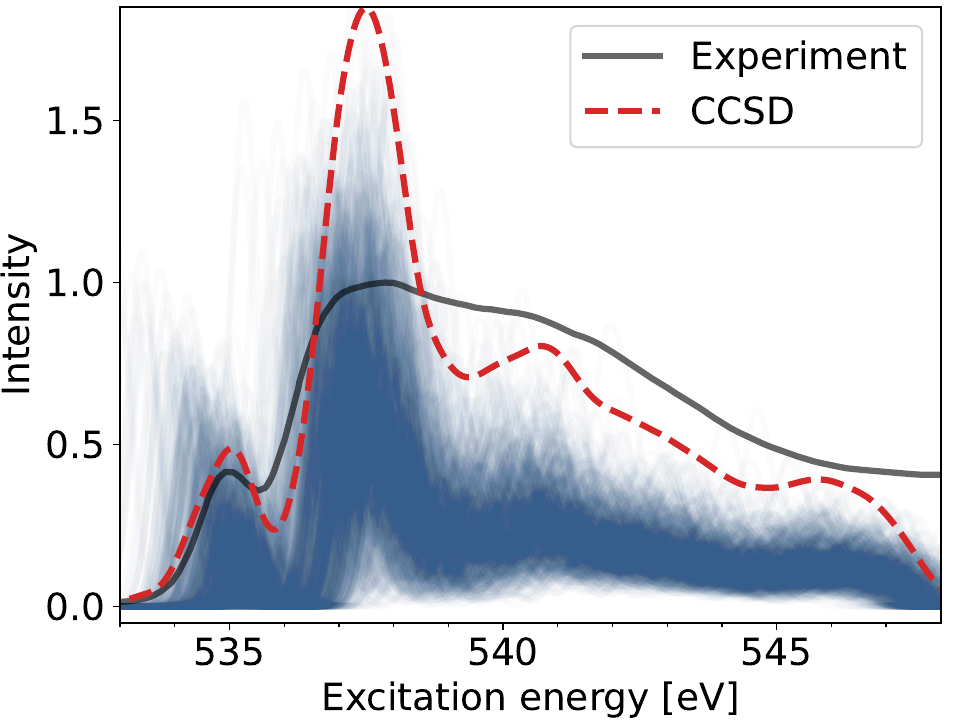}
    \end{subfigure}
    \begin{subfigure}[b]{0.49\textwidth}
        \includegraphics[width=\linewidth]{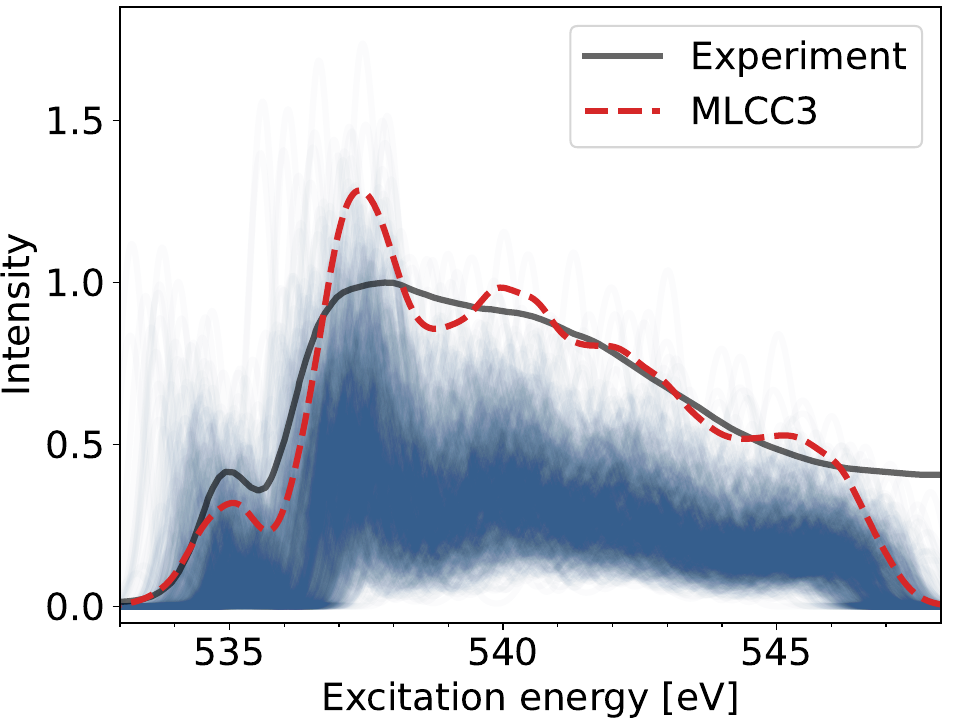}
    \end{subfigure}
    \caption{
        XAS spectra of water clusters at the CCSD-in-HF (left) and MLCC3-in-HF (right) level of  theory.
        The orange lines show the spectra averaged over 896 individual snapshots (blue)
        The calculated spectra were normalized to the experimental spectrum by
        matching the area under the curves in the range of $\SI{533}{}-\SI{545}{eV}$
        and a \textit{Gaussian broadening} with FWHM of $\SI{0.6}{\eV}$ was used.
        The CCSD-in-HF spectrum was shifted by $\SI{-1.5}{\eV}$ to match the experiment.
    }
    \label{App-fig:snapshots_gauss}
\end{figure}

\FloatBarrier

\section{Supplementary Information}\label{SI}
\subsection{Charge transfer analysis}
Here we describe the analysis of the charge transfer character
of the excitations in the X-ray absorption (XA) spectrum of liquid water
performed in this study.
Such type of analysis has been reported before
for TDDFT using Mulliken population analysis
in Ref.~\citenum{ChargeTransferNumbers0} and
more generally in Ref.~\citenum{ChargeTransferNumbers1}.

The change in density between an
excited state, $m$, and the ground state, $0$,
is related to the rearrangement of electrons due to their interaction with light.
This change in density can be calculated as the \textit{difference density},
\begin{equation}
    \DDM_m = \DM_m - \DM_0.
\end{equation}
As the trace of the one-electron density for an $N$-electron system
corresponds to the number of electrons,
the trace of the difference density must be zero.

\begin{figure}
    \centering
    \includegraphics[width=0.5\linewidth]{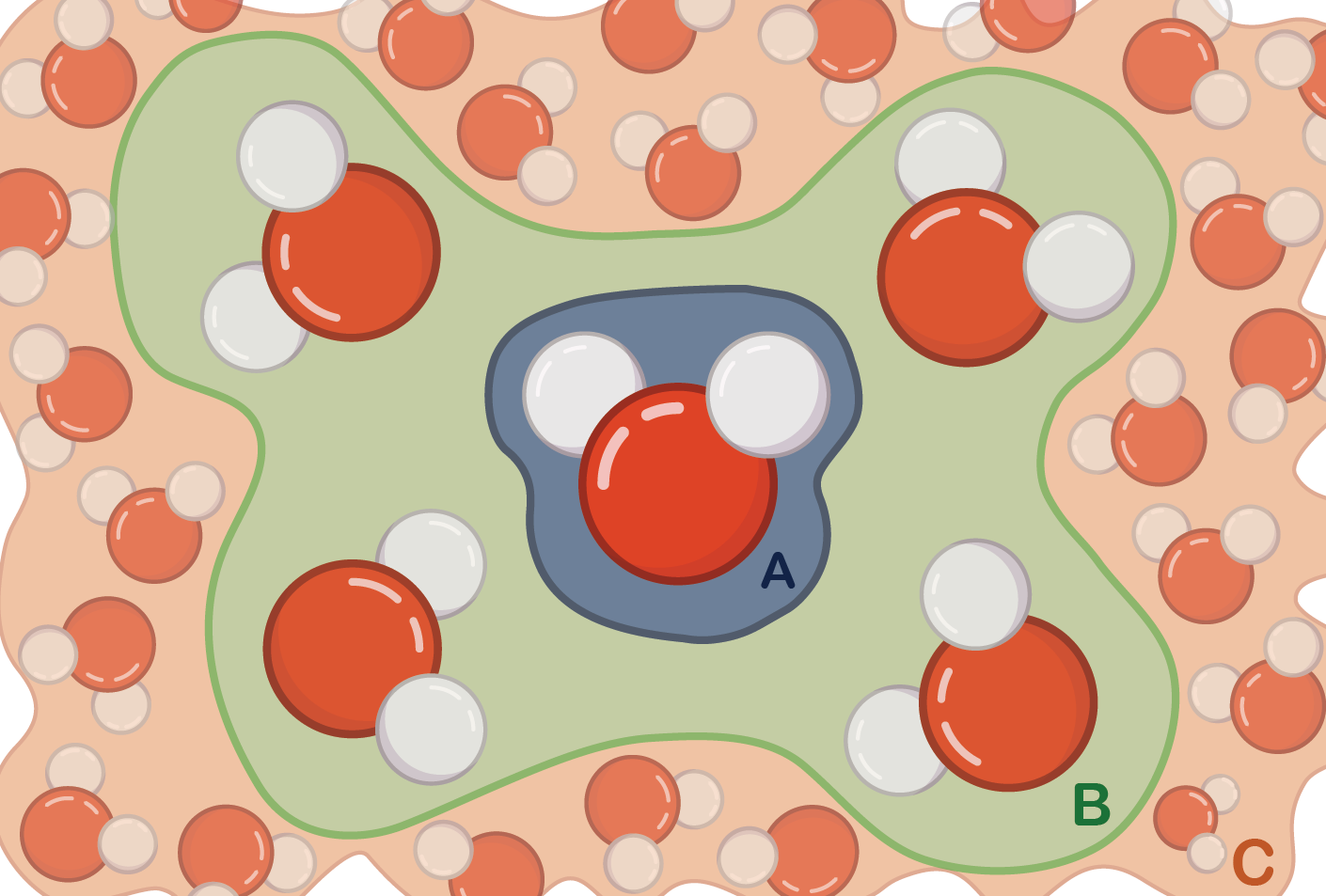}
    \caption{
        Illustration of the fragments used in the charge transfer analysis.
        \colorA{Fragment \sffamily{A}} contains only the central water molecule,
        \colorB{fragment \sffamily{B}} the four closest surrounding water molecules,
        and \colorC{fragment \sffamily{C}} the remaining water molecules which are treated at the HF level of theory.
    }
    \label{SI-fig:visualization_density_analysis}
\end{figure}

Using, for example, L{\"o}wdin population analysis,
we can transform the difference density
from the delocalized molecular orbital (MO) basis to the local atomic orbital (AO) basis,
\begin{equation}
    \DDM^{\textrm{AO}}_m = \SM^{\frac{1}{2}} \, \CM \, \DDM^{\textrm{MO}}_m \, \CM^T \, \SM^{\frac{1}{2}},
    \label{SI-eq:Loewdin-MO-AO}
\end{equation}
where $\SM$ is the atomic orbital overlap matrix and $\CM$ the matrix
containing the MO-coefficients.
Due to the cyclic property of the trace, the trace of the difference density in the AO basis is zero.
We can now partition the trace of the entire system into the trace of the three subsystems:
contributions from the central water molecule
(\colorA{Fragment \sffamily{A}} in Figure \ref{SI-fig:visualization_density_analysis}),
the four closest surrounding water molecules
(\colorB{Fragment \sffamily{B}} in Figure \ref{SI-fig:visualization_density_analysis}),
and the remaining water molecules
(\colorC{Fragment \sffamily{C}} in Figure \ref{SI-fig:visualization_density_analysis}),
\begin{equation}
    \tr_{\textrm{AO}} \DDM^{\textrm{AO}}_m
    = \colorA{\tr_{\textrm{AO} \in \textrm{\sffamily{A}}}} \DDM^{\textrm{AO}}_m
    + \colorB{\tr_{\textrm{AO} \in \textrm{\sffamily{B}}}} \DDM^{\textrm{AO}}_m
    + \colorC{\tr_{\textrm{AO} \in \textrm{\sffamily{C}}}} \DDM^{\textrm{AO}}_m
    = 0.
\end{equation}
The trace of a single subsystem is then interpreted
as the number of electrons detached from or attached to this subsystem,
depending on the sign.

\subsection{Active space approaches}
Because of the high cost of CCSD and CC3 prohibiting the treatment
of the entire system at the coupled cluster level,
we employ an active space approach in our study.
Two types of active spaces are used.
The first active space is constructed by localizing canonical Hartree-Fock orbitals for the target region.
Semi-localized Cholesky orbitals are used
for the occupied space and projected atomic orbitals (PAOs) for the virtual space.
The target region (active atoms) contains  the central water molecule
and the four closest surrounding water molecules of a given cluster,
as shown in Figure \ref{fig:method_illustration} of the main text.
The steps involved to construct these orbital spaces and to perform the MLCC3-in-HF calculation
are schematically illustrated in \ref{SI-fig:cc-in-hf-scheme}.

Cholesky orbitals are constructed by a restricted Cholesky decomposition
of the occupied Hartree–Fock density~\cite{folkestad2021multilevel},
using the atomic orbitals on the active atoms as pivots.
To ensure that the active density is localized,
diffuse orbitals (orbitals with an exponent smaller than one)
were excluded from the pivots.
Including diffuse functions will result in an active occupied orbital space that significantly exceeds the target region.
By excluding these orbitals from the pivots,
a more compact active space is obtained without
a significant loss in accuracy in the coupled cluster calculation.

For the virtual space,
the contribution of the occupied orbitals is projected out of
the atomic orbitals centered on the active atoms. These orbitals are the PAOs.
The final active virtual space is  obtained by removing linear dependencies and orthonormalizing the PAOs
using the L{\"o}wdin orthonormalization scheeme.

\begin{figure*}
    \centering
    \begin{subfigure}[b]{0.49\linewidth}
    \includegraphics[width=\linewidth]{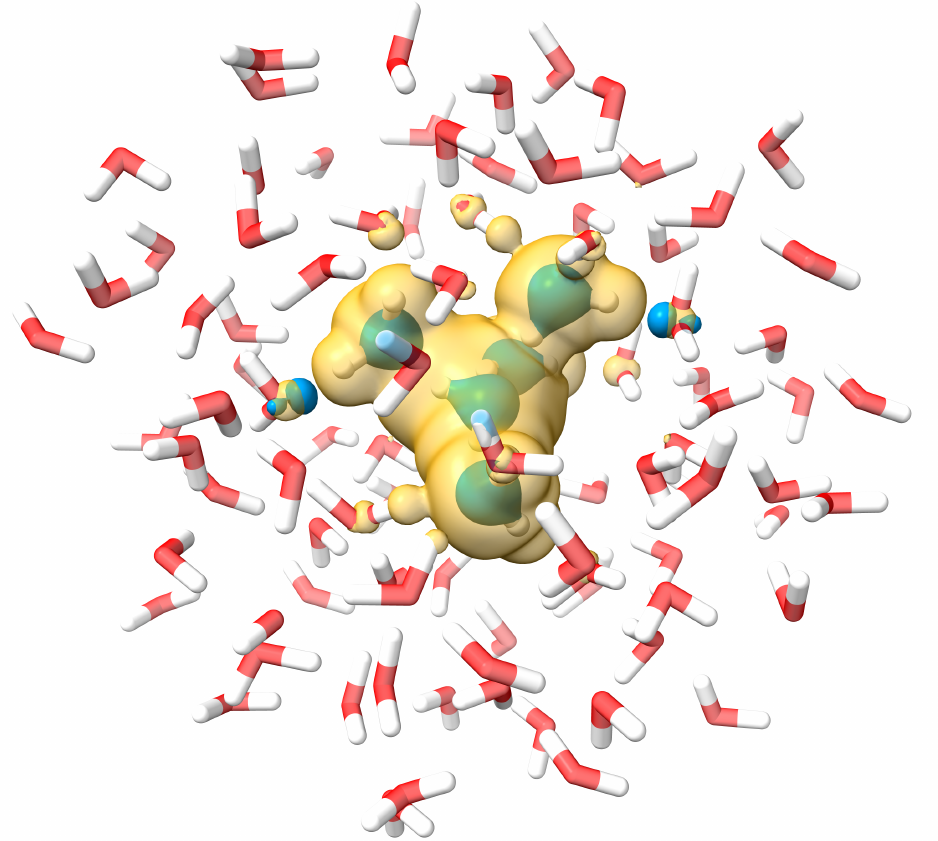}
    \end{subfigure}
    \begin{subfigure}[b]{0.49\textwidth}
    \includegraphics[width=\linewidth]{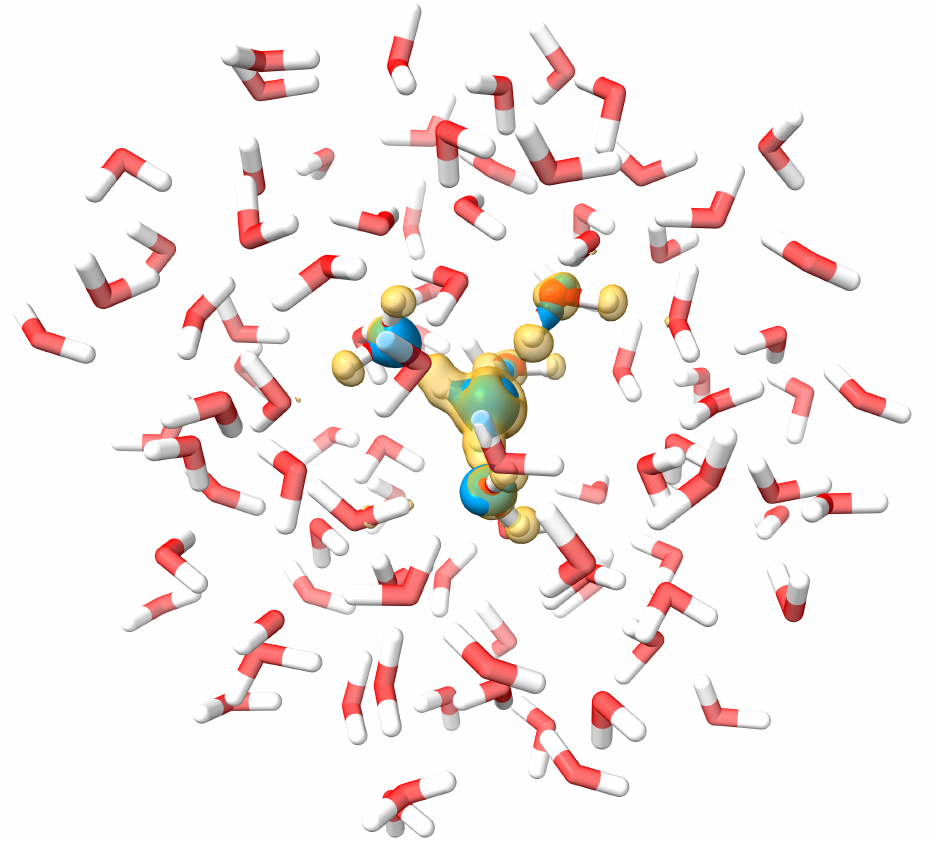}
    \end{subfigure}
    \caption{
        Left:
            Active occupied (blue) and virtual (yellow) CCSD-in-HF orbital densities,
            calculated from the localized MOs only.
        Right:
            Active occupied (blue) and virtual (yellow) MLCC3-in-HF orbital densities,
            calculated from the CNTOs only.
        For both plots an iso value of \SI{0.1}{\isovalue} was used.
    }
    \label{SI-fig:active_o_and_v_densities}
\end{figure*}

\begin{figure}
    \centering
    \includegraphics[width=0.8\linewidth]{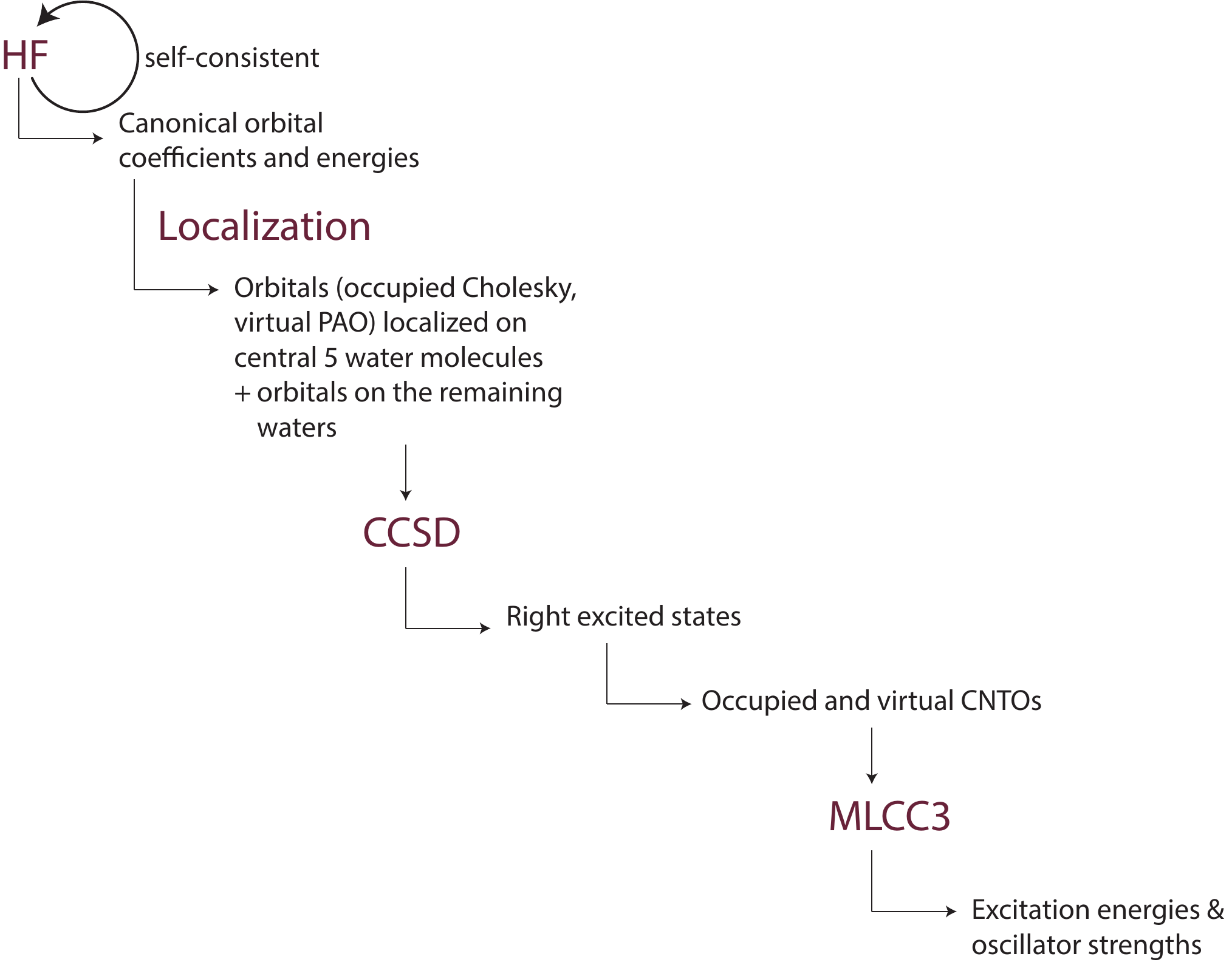}
    \caption{Illustration showing the computational workflow to perform a MLCC3-in-HF calculation.}
    \label{SI-fig:cc-in-hf-scheme}
\end{figure}

From this procedure, 21 to 24 (average over all snapshots 21.4)
occupied orbitals and 250 to 251 (average over all snapshots 251.0)
virtual orbitals were obtained.
This orbital space is comparable in size to a system consisting of
five water molecules, where one molecule is described using the aug-cc-pVTZ basis,
and the remaining four molecules with aug-cc-pVDZ and the frozen core approximation. 
In the left panel of Figure \ref{SI-fig:active_o_and_v_densities},
we depict the active occupied and virtual orbital densities which show that the orbitals
are largely localized on the five central water molecules.
The cluster operator in equation (1) of the main text is restricted to this active orbital space.
The inactive orbitals contribute to the calculation,
through a frozen Hartree-Fock density.
If the cluster operator is truncated after single and double excitations,
this approach is called CCSD-in-HF.

Since CC3 is more expensive than CCSD ($\mathcal{O}(N^7)$ versus $\mathcal{O}(N^6)$),
another active space is introduced to reduce the cost of the CC3 calculations.
This approach is called MLCC3,
where the triple excitations are restricted to this second active space,
while single and double excitations are included for the orbitals in the first active space,
see Figure 1 of the main text.
We use correlated natural transition orbitals (CNTOs) for the active space
in which we define the triple excitations~\cite{hoyvik2017correlated,folkestad2019multilevel}.
These orbitals have been shown to yield excellent results for core excitations with MLCC3~\cite{paul2022oscillator},
because they explicitly target the excited states of interest.
CNTOs are constructed by transforming the occupied and virtual orbitals with the eigenvectors of the matrices
$\boldsymbol{M}$ and $\boldsymbol{N}$,
respectively.
These matrices are defined as
\begin{align}
    \label{SI-eq:M}
   M_{ij} = \sum_a R^a_i R^a_j + \frac{1}{2} \sum_{abk}
   (1 + \delta_{ai,bk} \delta_{ij}) R^{ab}_{ik} R^{ab}_{jk}\\
   \label{SI-eq:N}
   N_{ab} = \sum_i R^a_i R^b_i + \frac{1}{2} \sum_{cij}
   (1+\delta_{ai,cj}\delta_{ab}) R^{ac}_{ij} R^{bc}_{ij},
\end{align}
where $\boldsymbol{R}$ is a CCSD excitation vector.
The magnitude of the eigenvalues of $\boldsymbol{M}$ and $\boldsymbol{N}$ determines how important a CNTO is
for the description of the excited state $\boldsymbol{R}$.
To ensure a well-balanced representation of all requested excited states,
we average the $\boldsymbol{M}$ and $\boldsymbol{N}$ over these states.
For the calculations performed in this study,
we computed 45 states and used the 12 most important occupied
and the 72 most important virtual CNTOs in the active space for MLCC3-in-HF.
The right panel in Figure \ref{SI-fig:active_o_and_v_densities}
shows the active occupied and virtual orbital density of the
CNTOs for a single snapshot.

\subsection{Difference densities visualization}
To visualize the importance of the environment for different parts
of the spectrum, we plot the difference densities for selected states
of a single snapshot, see Fig.\,\ref{App-fig:S68_step01_cc3_difference_34-1_and_4-1}
of \textit{Appendix \ref{Appendix}}.
The first excited state is largely localized on the central water.
Therefore,
the difference density between the first excited state and other excited states
shows the degree of delocalization of the higher excited state.
Comparing the density differences between excited states 1 and 4
(left panel of Fig.\,\ref{App-fig:S68_step01_cc3_difference_34-1_and_4-1})
and excited states 1 and 34
(right panel of Fig.\,\ref{App-fig:S68_step01_cc3_difference_34-1_and_4-1}),
we can see a gradual extension of the density from the central water to neighboring molecules.
This supports the results of the charge transfer analysis
that moving from the pre-edge to the post-edge,
the character of the excitation becomes more diffuse.

\subsection{Computational cost}
The calculations were performed on the \texttt{bigmem} nodes of the cluster \href{https://documentation.sigma2.no/hpc_machines/saga.html}{\textit{Saga}}.
These nodes consist of two Intel(R) Xeon(R) Gold 6138 CPU @ 2.00GHz with 20 threads each.
The calculations were run using 20 threads for parallelization and a local scratch directory for improved \texttt{I/O}.
Due to varying load on the nodes,
the timings fluctuate significantly for different inputs
which is reflected in the large standard deviation.

\begin{table}[ht]
    \centering
    \caption{Wall time and memory usage for the calculations run on \textit{Saga}.}
    \begin{tabular}{crr}
        \toprule
                        & CCSD                    & MLCC3 \\
        \midrule
        Wall time [h]  & $\SI{22.70 \pm 5.66}{}$ & $\SI{23.22 \pm 5.64}{}$ \\
        Memory [GB]    & $\SI{59.29 \pm 3.22}{}$ & $\SI{62.00 \pm 0.67}{}$ \\
        \bottomrule
    \end{tabular}
    \label{SI-tab:timings_Saga}
\end{table}

Note that the construction of the CNTOs in MLCC3 requires the CCSD right excited state vectors.
The time to construct the CCSD excited state vectors is, however, not included in the MLCC3 timings
because we restarted the MLCC3 calculations from the CCSD results.

\end{appendices}

\bibliography{paper}

\end{document}